\documentstyle[12pt,eqsecnum,aps,epsf]{revtex}

\tighten

\begin{document}

\def\lpmb#1{\mbox{\boldmath$#1$}}

\title{Computational Study of Baryon Number Violation \\
in High Energy Electroweak Collisions
}

\author{ Claudio Rebbi\footnote[1]{rebbi@pthind.bu.edu} and
Robert Singleton, Jr.\footnote[2]{bobs@cthulu.bu.edu}}
\smallskip
\address{Physics Department\\
Boston University\\
590 Commonwealth Avenue\\
Boston, MA 02215, USA
}
\maketitle

\setcounter{page}{0}
\thispagestyle{empty}

\vfill

\centerline{\bf Abstract}
\vskip0.2in

We use semiclassical methods to study processes which give rise to
change of topology and therefore to baryon number violation in the
standard model.  We consider classically allowed processes,
i.e.~energies above the sphaleron barrier.  We develop a computational
procedure that allows us to solve the Yang Mills equations of motion
for spherically symmetric configurations and to identify the particle
numbers of the in- and out-states.  A stochastic sampling technique is
then used to map the region spanned by the topology changing
solutions in the energy versus incoming particle number plane and, in
particular, to determine its lower boundary.  A lower boundary which
approaches small particle number would be a strong indication that
baryon number violation would occur in high energy collisions, whereas
a lower asymptote at large particle number would be evidence of the
contrary.  With our method and the computational resources we have had
at our disposal, we have been able to determine the lower boundary up
to energies approximately equal to one and a half time times the
sphaleron energy and observed a 40\% decrease in particle number
with no sign of the particle number leveling off.  However
encouraging this may be, the decrease in incoming particle
number is only from 50 particles down to approximately 30.
Nevertheless, the formalism we have established will make it
possible to extend the scope of this investigation and also to
study processes in the classically forbidden region, which we
plan to do in the future.

\vfill

\noindent BUHEP-95-33 \hfill Submitted to {\it Physical Review} {\bf D}

\noindent hep-ph/9601260 \hfill Typeset in REV\TeX
\eject

\baselineskip 24pt plus 2pt minus 2pt

\section{Introduction}
\hspace{1cm}
Since the pioneering work of \, 't Hooft\cite{thooft76} it has been
known that the axial vector anomaly implies that baryon number
is not conserved in processes which change the topology of the
gauge fields. Baryon number violating amplitudes are
non-perturbative and viable methods of calculation are scarce.
The two primary methods of obtaining non-perturbative
information in quantum field theory are either semi-classical
techniques or direct lattice simulations of the quantum
fluctuations. Theories with small coupling constants are not
suited for the latter, so the electroweak sector of the standard
model lies beyond the reach of direct lattice calculations.
This means that semiclassical methods presently offer the
only way to study baryon number violating electroweak
processes.

Electroweak baryon number violation is associated with topology
change of the gauge fields. Classically, gauge field configurations
with different topology (i.e.~differing by a topologically non-trivial
gauge transformation) are separated by an energy barrier. The
(unstable) static solution of the classical equations of motion
which lies at the top of the energy barrier is called the
sphaleron\cite{KM}. At energies lower than the sphaleron energy,
topology changing transitions, and hence baryon number violation,
can only occur via quantum mechanical tunneling. At zero
temperature and low energy the tunneling rate can be reliably
calculated and is exponentially small. A few years ago, however,
Ringwald \cite{ring} and Espinosa\cite{esp} noticed that a
summation of the semiclassical amplitudes over final states gives
rise to factors which increase very rapidly with increasing
energy. This may lead to a compensation of the exponential
suppression for energies approaching the energy of the
barrier, i.e.~the sphaleron energy $E_{\rm sph}$. Intuitively,
one might expect suppression of tunneling to become much
less severe as the energy approaches the energy of the barrier,
in particular, one might expect it to disappear altogether for
$E>E_{\rm sph}$, i.e.~in the region where the topology changing
processes are classically allowed. Investigations have indeed
confirmed that this is precisely what happens in high temperature
electroweak processes\cite{highT}: as the temperature
approaches $E_{\rm sph}$ (which is in fact temperature
dependent for a thermal plasma), the barrier-penetration
suppression factor becomes progressively less pronounced,
and electroweak baryon number violation becomes
unsuppressed altogether above the critical temperature.
The situation is, however, much less clear for high energy
collisions and it would be premature to conclude that
baryon number violation can occur with a non-negligible
amplitude. Phase space considerations are more subtle and
simply because one has enough energy to pass over the
barrier does not guarantee that one does so. The problem
is that in high energy collisions the incident state is an
exclusive two particle state, which is difficult to incorporate
in a semiclassical treatment of the transition amplitude.

A possible remedy to this situation has recently been proposed
by Rubakov, Son and Tinyakov\cite{rst} who suggested
that one considers incident coherent states, but
constrained so that energy and particle number take fixed
average values
\begin{mathletters}
\label{ENgsq}
\begin{eqnarray}
  E &=& {\epsilon \over g^2} \label{Egsq} \\
  N &=& {\nu \over g^2} .  \label{Ngsq}
\end{eqnarray}
\end{mathletters} \hskip -3pt
In the limit $g \to 0$, with $\epsilon$ and $\nu$ held fixed, the
path integrals giving the transition amplitudes are then dominated
by a saddle point configuration which solves the classical
equations of motion. This permits a semiclassical calculation
of the transition rates. Information on high energy collision
processes with small numbers of incident particles can then be
obtained from the limit $\nu \to 0$. While this limit does not
strictly reproduce the exclusive two-particle incoming state,
under some reasonable assumptions of continuity it can be
argued that the corresponding transition rates will be equally
suppressed or unsuppressed.

When the energy is below the sphaleron barrier the semiclassical
paths that dominate the functional integral in Ref.~\cite{rst}
must be complex for (\ref{ENgsq}) to be satisfied. Finding such
solutions is a formidable analytic problem, but one that is well
suited to numerical study. The numerical evolution naturally
divides into two regimes. There is a purely Euclidean evolution,
corresponding to tunneling under the barrier, and a Minkowski
evolution corresponding to classical motion before and after the
tunneling event. The desired semiclassical paths may be obtained
by appropriately matching the Euclidean and Minkowski solutions
onto one another, and the transition amplitude may then be
calculated.

When the energy is greater than the sphaleron barrier,
transitions are classically allowed and solutions that
saturate the functional integral are real. This is the regime
examined in this paper. When chiral fermions are coupled
to gauge and Higgs fields which undergo topological transitions,
Ref.~\cite{fggrs} shows that the anomalous fermion number
violation is given by the change in Higgs winding number of the
classical system. This paper is primarily an investigation of
whether and to what extent topology change occurs in classical
evolution with low particle number in the incident state. Since
Minkowski evolution is also required for the analysis below the
sphaleron, the techniques developed in the present investigation
will be useful there as well.

The primary impediment for rapid baryon number violation
is the phase space mismatch between incoming states of low
multiplicity and outgoing states of many particles. The authors
of Ref.~\cite{gr} look at simplified models and observe that,
classically, it is difficult to transfer energy from a small number
of hard modes to a large number of soft modes. However, the
investigations in Ref.~\cite{wdymt} find that for pure Yang-Mills
theory in \hbox{2-dimensions} the momenta can be dramatically
redistributed, although unfortunately the incident particle number
seems to be rather large in their domain of applicability.
Ref.~\cite{hmms} studies the Yang-Mills-Higgs system
in a \hbox{2-dimensional} plane-wave {\it Ansatz} and
again finds that momentum can be efficiently redistributed.
It is the purpose of our investigation to shed further light on
the situation in \hbox{4-dimensions} in the presence of a Higgs
field and to investigate the relation between incoming particle
number and topology change.

Given a typical classical solution, because of the dispersion of
the energy, the fields will asymptotically approach vacuum values.
Consequently, at sufficiently early and late times the field equations
will reduce to linearized equations describing small oscillations
about the vacuum and the field evolution will be a superposition
of normal mode oscillations. In terms of the frequencies
$\omega_n$ and amplitudes $a_n$ of these oscillators
the energy and particle number of (\ref{ENgsq}) are given by
\begin{mathletters}
\begin{eqnarray}
\epsilon &=& \sum_n \omega_n |a_n|^2
\label{esum} \\
\nu &=& \sum_n  |a_n|^2 \ ,
\label{nsum}
\end{eqnarray}
\end{mathletters} \hskip -5pt
and we see that for typical classical evolution the energy
$\epsilon$ and the particle numbers $\nu_i$ and $\nu_o$ of
the asymptotic incoming and outgoing states are well defined
(the energy is of course conserved and well defined even
in the non-linear regime, although no longer given by
(\ref{esum})). In addition, since the fields approach vacuum
values for $t \to \pm \infty$, the winding numbers of incoming
and outgoing configurations are also well defined. Because of
the sphaleron barrier, the energy $\epsilon$ of all the classical
solutions with a net change of winding number is bounded
below by the sphaleron energy $\epsilon_{\rm sph}$. The
problem we would like to solve then is whether the incoming
particle number $\nu_i$ of these solutions can be arbitrarily
small, or more generally, we would like to map the region
spanned by all possible values of $\epsilon$ and
$\nu_i$ for topology changing classical evolution.

One could easily parameterize an initial configuration of the
system consisting of incoming waves in the linear regime;
however, it would be extremely difficult to adjust the parameters
to insure that a change of winding number occurs in the course of
the subsequent evolution. For this reason we will instead
parameterize the configuration of the system at the moment when
a change of topology occurs (this will be our starting
configuration), and we will then evolve the equations of motion
backward in time. Following the time reversed evolution until
the system reaches the asymptotic linear regime allows us to
identify the incident particle number $\nu_i$. By varying the
parameters of the starting configuration with a suitable
stochastic procedure we will then be able to map the
boundary of the region of topology changing solutions
in the \hbox{$\epsilon$-$\nu$} plane.

Note that the problem of baryon number violation above the
barrier may roughly be divided into two parts. One must find
the set of incoming coherent states which give rise to a change
in topology of the fields, and one must calculate the overlap
between the incident two-particle scattering state and such
coherent states. Both are very challenging.
The problem considered in this paper is the more fundamental
of the two, in the sense that if topology change cannot occur
for coherent states with small average particle number,
the overlap effect with a two particle beam is a moot point.
On the other hand, if a change of topology can be induced
with arbitrarily low particle number in the incoming state,
one is at the very least assured that exponential suppression,
which is a residual of the barrier penetration, will be absent.

In summary, then, our strategy is the following. We start
with a (not necessarily small) perturbation about the
sphaleron with some energy $\epsilon$. We evolve the
configuration until it reaches the linear regime, at which
time we extract the normal mode amplitudes $a_n$ and
compute the asymptotic particle number $\nu$. The time
reversed solution will have an incident particle number
$\nu$ and will typically undergo topology change, since by
construction it will pass over the sphaleron barrier.
There is of course the possibility that the system will go
back over the sphaleron barrier and return to the original
topological sector, but we check against this occurrence
by evolving the starting configuration in the opposite
direction in time and measuring the winding number of the
asymptotic state. We can then explore the space of topology
changing solutions by varying the parameters of the starting
configuration using suitable stochastic techniques. This permits
us to map the allowed \hbox{$\epsilon$-$\nu$} plane in an attempt
to place a reliable lower bound on the incident particle number.
If this bound is comparable with two particles in the
incoming state, it would be an indication that the time
reversed solution, which passes over the sphaleron
barrier, can be excited in a high energy collision.
Hence, this would be a signal that baryon number
violation becomes unsuppressed. Likewise, if the
bound is large this would indicate that high energy
baryon number violation is unobservable in a two-particle
scattering experiment.

In what follows we put meat on the bones of the above
discussion and present our numerical results. The
structure of this paper is as follows. In Section~II we
illustrate the general properties of sequences of topology
changing field configurations, not necessarily
solutions to the equations of motion. For simplicity we
first consider the \hbox{2-dimensional} Abelian Higgs
model. We then examine the \hbox{4-dimensional} $SU(2)$
Higgs model, but restricted to the spherical
{\it Ansatz} to obtain a computationally tractable system.
In Section~III we examine the classical evolution in the continuum.
Since the field equations are coupled non-linear partial differential
equations, in Section~IV we solve them by numerical techniques.
In Section~V we describe the starting configurations
at the moment of topology change, i.e.~our parameterization
of the initial state, and in Section~VI we solve
the normal mode problem necessary for extracting
the particle number in the linear regime. In Section~VII we
explain the stochastic sampling technique used to probe
the initial configuration space and we present our
numerical results concerning the region spanned in the
\hbox{$\epsilon$-$\nu$} plane by topology changing
solutions. In Section~VIII we present concluding remarks
and directions for future research. The reader who is familiar
with the the basic properties of the $SU(2)$ Higgs system
and of topology changing solutions, and is impatient to
learn about our results, may skip directly to Section~VII.
However, in our opinion, much of the value of the research
we present here is to be found in the formalism
we have established to parameterize, evolve and analyze
classical solutions of the $SU(2)$ Higgs system in
the spherical {\it Ansatz}. This formalism, which
is illustrated in Sections~II through~VI, has not
only been crucial for obtaining our current results,
but we are confident it will be invaluable for
further investigation into the problem of
collision-induced baryon number violation
both above and below the sphaleron barrier.

\section{Topology Changing Sequences of Configurations}
\hspace{1cm}
We start our investigation with
the \hbox{1+1} dimensional Abelian Higgs system, which
is defined
in terms of a complex scalar field $\phi(x)$ and an
Abelian gauge potential $A_{\mu}(x)$ with action
\begin{equation}
S = \int dx ^2 ~ \left\{- {1 \over 4} F_{\mu \nu} F^{\mu \nu} +
D_{\mu}  \phi^* D^{\mu} \phi  - \lambda (|\phi|^2 -1 )^2
\right\} \ ,
\label{eq4}
\end{equation}
where the indices run over 0 and 1, $F_{\mu \nu} =
\partial_{\mu} A_{\nu} - \partial_{\nu} A_{\mu}$ and
$D_{\mu}\phi = \partial_{\mu}\phi - i A_{\mu} \phi$. We have
set the coupling constant $g=1$ and several inessential
constants have been eliminated by a suitable choice of units.

The most important feature of this system is that the vacuum,
i.e.~the configuration of minimum energy, occurs for
non-vanishing $\phi$, indeed, in our units for $|\phi| =1$.
Since this does not specify the phase of $\phi$, there is not
a unique vacuum state, but rather multiple vacua. Still,
because of gauge invariance one must be careful in regard
to the physical significance of the phase of $\phi$. A local
variation in the phase of $\phi$ can always be undone by
a suitable gauge transformation, and since gauge
equivalent configurations must be considered physically
indistinguishable, local variations of the phase of the scalar
field do not lead to different vacua. However, variations
of the phase of $\phi$ by multiples of $2 \pi$ (as the coordinate
$x^1$ spans the entire spatial axis) cannot be undone by a
local gauge transformation, and thus define topologically
distinct vacuum states. These vacua differ by the global
topological properties of the field configuration. The condition
$|\phi|=1$ restricts the values of the scalar field to the unit
circle (in the complex plane). In the $A_0=0$ gauge, which
we use throughout this paper, the values assumed by $\phi$
at $x^1=\pm\infty$ stay constant in time. If we demand that
$\phi$ takes fixed identical values as $x^1 \to \pm \infty$
(a condition we later relax), then the number of times $\phi$
winds around the unit circle as $x^1$ spans the entire real
axis is a topological invariant (the winding number) which
characterizes different topologically inequivalent vacuum
states.

\vbox{
\begin{figure}
\centerline{
\epsfxsize=120mm
\epsfbox{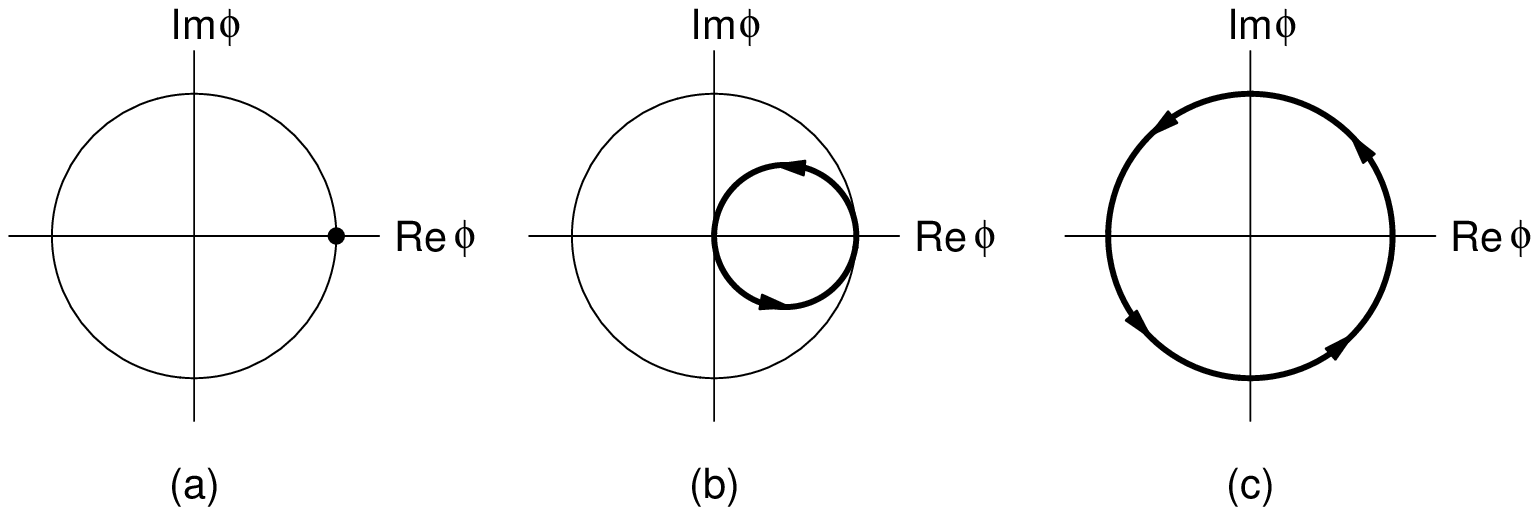}
}
\centerline{
\epsfxsize=135mm
\epsfbox{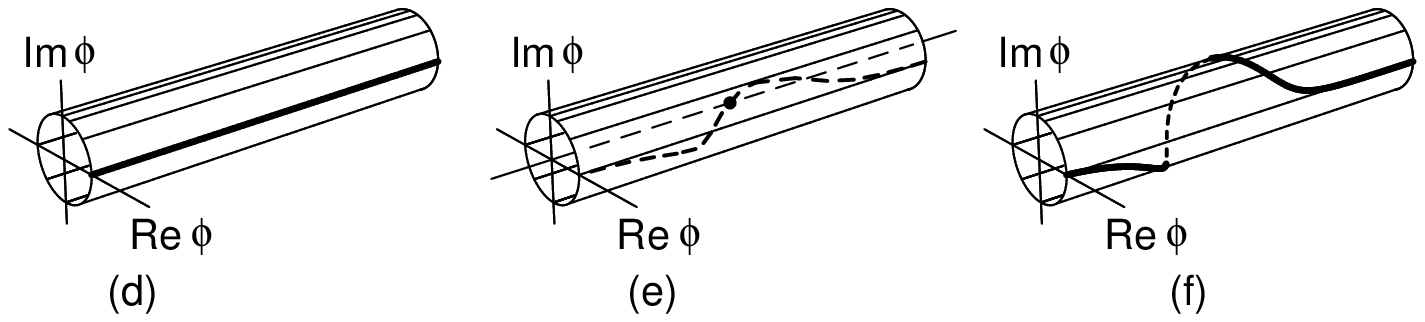}
}
\caption{\tenrm \protect\baselineskip 12pt
Example of two inequivalent vacuum configurations (a, c)
and a field configuration at the top of the energy barrier
separating them (b). Figures a, b and c trace the field
$\phi$ in the
complex plane as the spatial coordinate spans the entire
axis. A three dimensional perspective has been added
in figures d, e and f to illustrate the detailed dependence
of $\phi$ on the spatial coordinate.
}
\end{figure}
}

Figures 1a-c illustrate three possible contours traced in
the complex plane by the field variable $\phi(x^1)$ as the
coordinate $x^1$ spans the entire space axis. Inequivalent
vacuum configurations with winding numbers 0 and 1
respectively are depicted in Figs.~1a and 1c. In the contour
of Fig.~1a the phase of $\phi$ stays fixed at zero as $x^1$
ranges between $-\infty$ and $+\infty$, whereas it goes
once around the unit circle in Fig.~1c. Consequently, the
corresponding vacuum configurations have winding
numbers $0$ and $1$. The detailed variation
of the phase is immaterial since it can always be changed
locally by a gauge transformation. Thus, in Fig.~1a for
example, as $x$ varies from $-\infty$ to $+\infty$ the field
does not have to stay fixed, but could wander continuously
on the unit circle provided the net change in phase is zero.
However, the configuration of Fig.~1a cannot be continuously
deformed to that of Fig.~1c without leaving the vacuum manifold.
Therefore a continuous path of configurations connecting
neighboring vacua must pass over an energy barrier, a
configuration which has the property that $\phi$ vanishes
at a point, rendering its phase there undefined. The smallest
such energy barrier is called the sphaleron\cite{KM}, and
its Higgs field component is illustrated in Fig.~1b. Figures ~1d-f
add the additional perspective of spatial dependence for the
field $\phi(x^1)$. Figures~1a-c can be viewed as projections
onto the complex plane orthogonal to the $x^1$ axis of the
curves in Figs.~1d-e.

\vbox{
\begin{figure}
\centerline{
\epsfxsize=120mm
\epsfbox{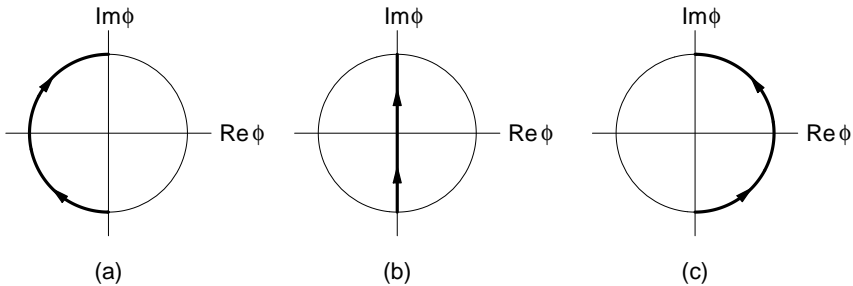}
}
\caption{\tenrm
A different gauge equivalent representation of
the configurations illustrated in Fig.~1.
}
\end{figure}
}

One should note that the periodic boundary conditions
on $\phi$ at $x^1 = \pm \infty$ can be relaxed. Sometimes
it is convenient to use the freedom of performing a time
independent gauge transformation to make $\phi(\infty)$
and $\phi(-\infty)$ differ while keeping both fixed in time
(for solutions, the constancy in time of $\phi(\pm \infty)$
follows from the equations of motion in the $A_0=0$ gauge).
Thus, the configurations of Figs.~1a-c can be gauge
transformed into the configurations shown in Figs.~2a-c.
In Fig.~2a the phase of $\phi$ changes by $-\pi$ as $x^1$
goes from $-\infty$ to $+\infty$, while in Fig.~2c it rotates
by $\pi$. As in Fig.~1, the two vacuum configurations
differ by a phase rotation of $2\pi$, i.e.~by a unit change
of winding number. In the intermediate configuration
(Fig.~2b) the scalar field takes only imaginary values. In this
gauge the sphaleron configuration takes a very simple form
\begin{equation}
\phi(x^1)=i \tanh [\sqrt{\lambda}\, (x^1-c)], \quad A_{\mu}=0,
\label{phitanh}
\end{equation}
where $c$ specifies the location of the sphaleron.

A possible parameterization for the entire evolution illustrated
in Fig.~2 can be conveniently written as
\begin{mathletters}
\label{phiapath}
\begin{eqnarray}
\phi(x^1)&=&i {1-\exp [ i \tau -2 \sqrt{\lambda}\, (x^1-c)] \over
1+\exp [i \tau -2 \sqrt{\lambda}\, (x^1-c)] } \\
A_1&=&
{4 \tau \sqrt{\lambda} \over \pi \cosh [2 \sqrt{\lambda} \,
(x^1-c)]} \ ,
\label{apath}
\end{eqnarray}
\end{mathletters} \hskip -3pt
with $A_0=0$.
As the reader can easily verify, for $\tau=-\pi/2$ and
$\tau=\pi/2$ the field $\phi$ reduces to a number of
unit modulus precisely spanning the contours of Fig.~2a
and Fig.~2c respectively (as $x^1$ ranges from $-\infty$
to $+\infty$). The corresponding values of $A_1$ are
chosen to make the gauge covariant derivative of $\phi$
vanish, thus ensuring vacuum.  We should point out, however,
that (\ref{phiapath}) does not represent the solution of
any particular set of equations of motion (Euclidean or Minkowski).
It is merely a compact parameterization of interpolating
configurations, in terms of two variables $c$ and $\tau$, which
might be useful in studying sphaleron transitions based on the
method of collective coordinates.

Classical solutions of the \hbox{2-dimensional} Abelian Higgs
model can exhibit topology change in much the same way as the
vacuum-to-vacuum paths described above. If one couples
chiral fermions to the system, the fermionic current has an
anomaly which leads to fermion number violation in the
presence of topology changing classical solutions. Therefore,
this model would appear to be a very convenient system
for a simplified study of baryon number violation in high
energy processes. However, as we will discuss in a future
section, a crucial component of the computational
investigation is the ability to identify numerically the
normal mode amplitudes of the fields in the asymptotic
linear regime. No matter how non-linear the system may be at
any given point in its classical evolution, typically
the energy will disperse and bring the system to a regime
where the fields undergo small oscillations about a vacuum
configuration. This dispersion is expected to occur in any
field theoretical system, unless prevented by conservation
laws such as those underlying soliton phenomena. Now, while
the \hbox{2-dimensional} Abelian Higgs model does
not possess soliton solutions, we have observed computationally
that the decay of the sphaleron in this system nevertheless
gives origin to persistent, localized, large oscillations with
an extremely small damping rate (this observation was also
made by Arnold and McLerran in Ref.~\cite{am88}). These
oscillations, illustrated in Fig.~3, make the system quite
unwieldy for a computational investigation of baryon number
violation based on semiclassical techniques. Consequently
we turn our attention to the more realistic \hbox{4-dimensional}
$SU(2)$ Higgs system.

\vbox{
\begin{figure}[b]
\centerline{
\epsfxsize=95mm
\epsfbox{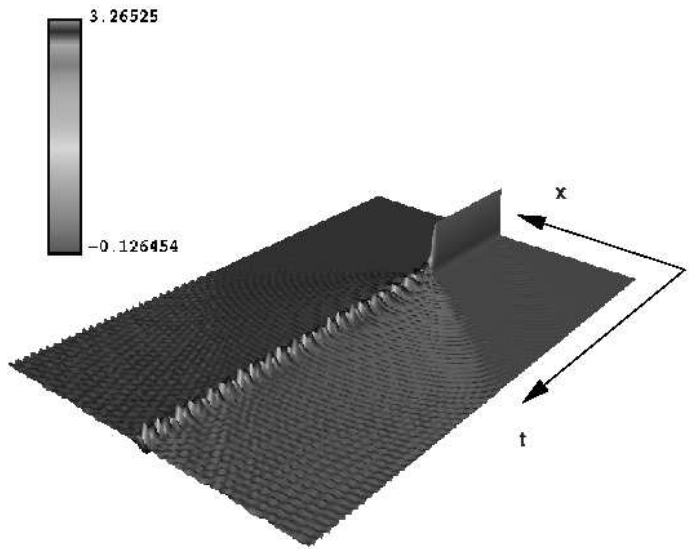}
}
\caption{\tenrm
Sphaleron decay in the \hbox{2-dimensional} Abelian Higgs model:
classical evolution of the $\phi$ field. The values of the phase
of the complex field are coded by shades of gray, and the modulus
of the field by the height of the surface. The sphaleron decays
rather quickly, but leaves behind a quasi-stable oscillating
remnant. For a full color figure see
http://cthulu.bu.edu/$\scriptstyle\sim$bobs/bviolate.html.
}
\end{figure}
}
\vskip1.5in

Throughout this paper we will ignore both the $U(1)$ hypercharge
and the back-reaction of the fermions on the dynamics of the
gauge and Higgs fields. We shall examine the \hbox{3+1}
dimensional $SU(2)$ Higgs system, which is defined in terms of a
complex doublet $\Phi(x)$ and a gauge potential $A_\mu(x)$ with
action
\begin{equation}
S = \int dx ^4 ~ \left\{- {1 \over 2} {\rm Tr}\,F_{\mu \nu} F^{\mu \nu} +
(D_{\mu} \Phi)^\dagger D^{\mu} \Phi  - \lambda (\Phi^\dagger
\Phi -1 )^2
\right\} \ ,
\label{fourAction}
\end{equation}
where the indices run from $0$ to $3$ and where
\begin{eqnarray}
F_{\mu\nu} &=& \partial_\mu A_\nu -  \partial_\nu A_\mu
 - i [A_\mu,A_\nu] \\
D_\mu \Phi &=&  (\partial_\mu - i A_\mu) \Phi
\end{eqnarray}
with $A_\mu = A^a_\mu\sigma^a/2$. We use the standard metric
$\eta_{\mu\nu}={\rm diag}(1,-1,-1,-1)$, and we have eliminated
several inessential constants by a suitable choice of units. We
have also set the coupling constant $g=1$, but shall restore
it when explicitly needed using the standard model value
$g=0.652$. For our numerical investigation we shall take the
Higgs self-coupling $\lambda=0.1$, which corresponds to
$m_{\rm H}= 110 {~\rm GeV}$. This value of $\lambda$ is small
enough that Higgs-field dynamics is non-trivial, but large enough
to allow many lattice sites to fall within a single Higgs Compton
wave length.

Because of the larger dimensionality of space one expects the
energy to disperse much more readily in this system than in the
1+1 dimensional Abelian Higgs model, an expectation borne out by
results of Hellmund and Kripfganz\cite{hk91} who observed the
onset of a linear regime following the sphaleron's decay. For
a computationally manageable problem, we focus on the spherically
symmetric configurations of Ratra and Yaffe\cite{ry88}, which
reduce the system to an effective \hbox{2-dimensional} theory.
This effective theory, however, still has much in common with
the full \hbox{4-dimensional} theory, such as possessing
similar topological structure. Furthermore, despite its lower
dimensionality, we shall see that the effective system still
linearizes because of explicit kinematic factors of $r$ in the
equations of motion (these factors are lacking for the 1+1
dimensional Abelian Higgs model). The ease of linearization in
this effective \hbox{2-dimensional} theory is physically
reasonable since solutions within the spherical ansatz can have
their energy distributed over expanding spherical shells.

Explicitly, the spherical {\it Ansatz} is given by expressing the
gauge and Higgs fields  in terms of six real functions $a_0\, ,\,a_1\, ,
\, \alpha\, , \, \beta\, , \, \mu\ {\rm and}\ \nu\ {\rm of}\ r\
{\rm and}\ t$:
\begin{mathletters}
  \label{SphAn}
\begin{eqnarray}
  A_0({\bf x},t) &=& \frac{1}{2 } \, a_0(r,t)
\lpmb\sigma \cdot {\bf\hat x}
  \label{sphao}\\
  A_i({\bf x},t) &=& \frac{1}{2 } \, \big[a_1(r,t)
\lpmb\sigma \cdot {\bf\hat x}
 \hat  x^i+\frac{\alpha(r,t)}{r}(\sigma^i- \lpmb\sigma
  \cdot {\bf\hat x}\hat x^i)
  +\frac{1+\beta(r,t)}{r}\epsilon^{ijk}\hat x^j\sigma^k\big]
\label{sphai}\\
  \Phi({\bf x},t) &=&   [ \mu(r,t) + i \nu(r,t)\lpmb\sigma
\cdot {\bf\hat x} ] \xi  \ ,
  \label{sphh}
  \end{eqnarray}
\end{mathletters} \hskip -4pt
where ${\bf \hat x}$ is the unit three-vector in the radial direction
and $\xi$ is an arbitrary two-component complex unit vector.
For the \hbox{4-dimensional} fields to be regular at the origin, $a_0$,
$\alpha$, $a_1 - \alpha/r$, $(1+\beta)/r$ and $\nu$ must vanish
like some appropriate power of $r$ as $r \to 0$.

Note that configurations in the spherical {\it Ansatz} remain
in the spherical {\it Ansatz} under gauge transformations of the
form
\begin{eqnarray}
 \label{sphgt}
 A_\mu &&\to A_\mu + i U^\dagger \partial_\mu U
 ~~~~~~ \mu=0\cdots3  \\
  \Phi \, &&\to U \Phi \ ,
\end{eqnarray}
where the gauge function is given by
\begin{equation}
\label{Usph}
U=\exp[i\Omega(r,t)\mbox{\boldmath$\sigma$}
\cdot {\bf\hat x}/2]  \ .
\end{equation}
We require $\Omega(0,t)=0$ to ensure that gauge transformed
configurations of regular fields remain regular at the origin.
This spherical gauge degree of freedom induces a residual
$U(1)$ gauge invariance in an effective \hbox{2-dimensional}
theory. The action of this effective theory can be obtained by
inserting (\ref{SphAn}) into (\ref{fourAction}), from which one
finds
  \begin{eqnarray}
  S =  4\pi \int dt\int^\infty_0dr  &&\bigg[-\frac{1}{4}
  r^2f^{\mu\nu}f_{\mu\nu}+D^\mu \chi^* D_\mu \chi
  + r^2 D^\mu\phi^* D_\mu\phi
  \nonumber\\
  && -\frac{1}{2 r^2}\left( ~ |\chi |^2-1\right)^2
  -\frac{1}{2}(|\chi|^2+1)|\phi|^2 -  {\rm Re}(i \chi^* \phi^2)
  \nonumber \\
  && -\lambda  \, r^2 \, \left(|\phi|^2 - 1\right)^2 ~
  \bigg] \ ,
  \label{effAction}
  \end{eqnarray}
where the indices now run from $0$ to $1$ and in contrast
to Ref.~\cite{ry88} are raised and lowered with
$\eta_{\mu\nu}={\rm diag}(1,-1)$, and where
\begin{eqnarray}
  f_{\mu\nu}&=& \partial_\mu a_\nu-\partial_\nu a_\mu\
\label{defConva}   \\
  \chi &=&\alpha+i \beta
\label{defConvb} \\
  \phi &=& \mu+i \nu\
\label{defConvc} \\
  D_\mu \chi &=& (\partial_\mu- i   \, a_\mu)\chi
\label{defConvd}  \\
  D_\mu \phi&=& (\partial_\mu - \frac{i}{2}  \, a_\mu)\phi\ .
\label{defConve}
\end{eqnarray}

The action (\ref{effAction}) is indeed invariant under
the $U(1)$ gauge transformation
\begin{mathletters}
\label{gaugexform}
\begin{eqnarray}
 a_\mu &&\to a_\mu + \partial_\mu \Omega
\label{amueq}\\
  \chi \, &&\to e^{i \Omega} \chi
\label{chieq}\\
  \phi \, &&\to e^{i \Omega/2} \phi  \ ,
\label{phieq}
\end{eqnarray}
\end{mathletters} \hskip -3pt
and we see that the spherical {\it Ansatz} effectively
yields a system very similar to the Abelian Higgs model
considered above. In this reduced system the variables
$a_0(r,t)$ and $a_1(r,t)$ play the role of the
\hbox{2-dimensional} gauge field. The variables $\chi(r,t)$
and $\phi(r,t)$, which parameterize the residual components of
the \hbox{4-dimensional} gauge field and the \hbox{4-dimensional}
Higgs field respectively, both behave as \hbox{2-dimensional}
Higgs fields. Note that $\chi$ has a $U(1)$ charge of one while
$\phi$ has charge one half. Of course, the presence of metric
factors (powers of $r$) in the action (\ref{effAction}) is a
reminder that we are really dealing with a \hbox{4-dimensional}
system.

We shall work in the $a_0=0$ (or $A_0=0$) gauge throughout.
In the \hbox{4-dimensional} theory, if one compactifies
\hbox{3-space} to $S^3$ by identifying the points at infinity, it
is well known that the vacua correspond to the topologically
inequivalent ways of mapping $S^3$ into $SU(2)\sim S^3$
\cite{JR}. These maps are characterized by the third homotopy
group of $SU(2)$ and a vacuum can be labeled by an integer
called the homotopy index or winding number. The effective
\hbox{2-dimensional} theory inherits a corresponding vacuum
structure. From (\ref{effAction}) it is apparent that the vacuum
states are characterized by $|\chi|=|\phi|=1$, with the
additional constraint that $i \chi^* \phi^2 =-1$ (as well as
$D_1\chi =D_1 \phi =0$). Convenient zero-winding vacua
are given by $\chi_{\rm vac}=-i$, $\phi_{\rm vac}=\pm 1$
with $a_{1 \, \rm vac}=0$. There are in fact other vacua with
constant fields (and hence zero winding), but from (\ref{SphAn})
they yield singular \hbox{4-dimensional} fields. Nontrivial
vacua can be obtained from the trivial vacua via the gauge
transformation
(\ref{gaugexform}):
\begin{mathletters}
\label{sphvac}
\begin{eqnarray}
 a_{\mu \, \rm vac} &=&  \partial_\mu \Omega
\label{amuvac}\\
  \chi_{\rm vac} \, &=& -i \, e^{i \Omega}
\label{chivac}\\
  \phi_{\rm vac} \, &=& \pm \, e^{i \Omega/2} \ .
\label{phivac}
\end{eqnarray}
\end{mathletters} \hskip -3pt
When \hbox{3-space} is compactified $\Omega \to 2 n \pi$
as $r \to \infty$ (for non-zero \hbox{integers $n$}). Since
$\Omega$ has been set to zero at the origin, the winding
numbers of such vacua are simply the integers $n$.
Note that $\chi_{\rm vac}$ winds $n$ times around the unit
circle while $\phi_{\rm vac}$ only winds by $n/2$. This is
because the $\phi$ field has half a unit of $U(1)$ charge while
$\chi$ has a full unit. Hence, the phase change of $\chi$ is
more dramatic in a topological transition, and for this reason
we will often concentrate our attention upon $\chi$ rather than
$\phi$, even though the Higgs field is more fundamental for
topology change \cite{fggrs}.

As will become apparent shortly, it is often convenient to
relax the condition that \hbox{3-space} be compactified. We
may then consider vacua (\ref{sphvac}) for which $\Omega$
does not become an even multiple of $\pi$ at large $r$. In
particular, when $\Omega \to (2 n + 1)\pi$, then $\chi_{\rm vac}
\to i$ and $\phi_{\rm vac} \to \pm \, i$ as $r \to \infty$. Then the
gauge function $U \to \pm \, i \lpmb\sigma \cdot {\bf \hat x}$
and becomes direction dependent, and as expected, space
cannot be compactified.

As in the Abelian Higgs model a continuous path in the space
of all field configurations which interpolates between two
inequivalent vacua must necessarily leave the manifold of
vacuum configurations and pass over an energy barrier. On
such a path there will be a configuration of maximum energy,
and of all these maximal energy configurations the sphaleron
has the lowest energy and represents a saddle point along the
energy ridge separating inequivalent vacua\cite{KM}. In the
spherical {\it Ansatz} we can work in a gauge in which the
sphaleron takes a particularly simple form, with $a_\mu=0$
and
\begin{eqnarray}
\label{sphSphal}
&&\chi_{\rm sph}(r)=i [2 f(r)-1] \\
\nonumber
&& \phi_{\rm sph}(r)=i h(r)  \ ,
\end{eqnarray}
where $f$ and $h$ vary between $0$ and $1$ as $r$ changes from
$0$ to $\infty$ and are chosen to minimize the energy functional.
Note that the $\phi$ field vanishes at the origin and that the
$\chi$ field vanishes at some non-zero value of $r$.

This form of the sphaleron, in which the gauge field $a_\mu$
vanishes and the fields $\chi$ and $\phi$ are pure imaginary,
is convenient for numerical calculations. Nevertheless, it is
slightly peculiar in the following sense. Finite energy
configurations, like (\ref{sphSphal}), asymptote to pure
gauge at spatial infinity (note that $i \chi^*_{\rm sph}
\phi^2_{\rm sph} \to -1$ as $r \to \infty$). Typically a gauge
is chosen so that the appropriate gauge function is unity at
spatial infinity, and then space can be compactified to the
\hbox{3-sphere}. But (\ref{sphSphal}) gives $\chi_{\rm sph}
\to i$ and $\phi_{\rm sph} \to i$, which as we have seen in
the discussion following (\ref{sphvac}) corresponds to
the direction dependent gauge function $U \to i \lpmb\sigma
\cdot {\bf \hat x}$. So the sphaleron (\ref{sphSphal}) is in a
gauge in which {3-space} cannot be compactified. Note that an
arbitrary element of $SU(2)$ can be parameterized by ${\rm b}_0
{\bf 1} + i \mbox{\boldmath$ \sigma$}\cdot {\bf b}$ where
${\bf 1}$ is the two by two unit matrix and ${\rm b}_0^2+{\bf
b}^2=1$. Hence $SU(2)\sim S^3$, and defining the north and south
poles by $\pm {\bf 1}$, we see that $i \mbox{\boldmath$\sigma$}
\cdot {\bf b}$ with ${\bf b}^2=1$ parameterizes the equatorial
sphere. Thus the gauge function $U$ maps the sphere at infinity
onto the equatorial sphere of $SU(2)$. In this gauge, a topology
changing transition proceeding over the sphaleron corresponds to
a transition where the fields wind over the lower hemisphere of
$SU(2)$ before the transition and over the upper hemisphere after
the transition, with a net change in winding number still equal
to one. The behavior of the $\chi$ field in a topological
transition is then very similar to the behavior of the Higgs
field in the \hbox{2-dimensional} model, already illustrated in
Fig.~2. The behavior of the $\phi$ field is illustrated in
Fig.~4. We could of course, and sometimes will, work in a gauge
consistent with spatial compactification where topological
transitions interpolate between vacua of definite winding, as in
Fig.~1, but the sphaleron would look more complicated. The
advantage of (\ref{sphSphal}) from a computational perspective
is that perturbations about the sphaleron can be more easily
parameterized.

\vbox{
\begin{figure}
\centerline{
\epsfxsize=120mm
\epsfbox{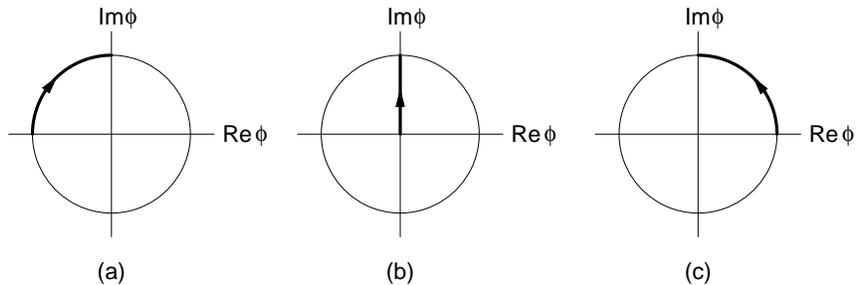}
}
\caption{\tenrm
Topological transition in the \hbox{4-dimensional}
$SU(2)$ Higgs model: behavior of the $\phi$ field.
The $\chi$ field behaves as in Fig.~2.
}
\end{figure}
}

\section{Classical Evolution in the Continuum}
\hspace{1cm}
So far we have only examined topology changing paths that
interpolate between inequivalent vacua. We are now interested
in examining the topological structure of solutions to the
equations of motion. For vacuum to vacuum sequences it is clear
what we mean by topology change: this is simply the change in
winding number between the initial and final vacua. For
solutions, however, the situation is not quite so straightforward.
Nevertheless, topology change can be precisely defined for
solutions whose energy density dissipates to zero uniformly
in the distant past and future, which is the generic case for
classical evolution. In the asymptotic regime the uniform
dissipation of energy renders the system linear and the
waves can be expressed as small oscillations about vacua
of definite winding numbers. By the topology change of such
a solution, we simply mean the difference in the winding
number between these two asymptotic vacua. This difference
in winding is in fact just given by the change in Higgs winding
number, and hence is characterized by zeros of the Higgs field
(although in the spherical {\it Ansatz} it is characterized by zeros
of both $\phi$ and $\chi$). The most important physical
consequence of this topology change is that when chiral
fermions are coupled to the system, fermion number violation
occurs and is proportional to the change in winding of the
Higgs field (see Ref.~\cite{fggrs}).

We wish to study whether topology change, and hence fermion
number violation, can occur in the course of classical evolution
with small gauge or Higgs particle-number in the incoming state.
Since the system we are studying linearizes in the past, the
incident particle number is defined and our question is well
posed. However, the field equations are coupled non-linear
partial differential equations which we cannot solve in closed
form. Our approach, then, is to solve the equations numerically
with a discretized $r$-axis and discretized time steps, but first
it is useful to examine the continuum system.

The equations of motion obtained from the action
(\ref{effAction}) are

\vbox{
\begin{mathletters}
  \begin{equation}
  \partial^\mu(r^2f_{\mu\nu})=i \left[D_\nu \chi^*\chi-\chi^*
  D_\nu\chi \right] + \frac{i }{2}\,   \, r^2 \left[D_\nu \phi^*\phi-
  \phi^*D_\nu\phi\right]
  \label{fEq}
  \end{equation}
  \begin{equation}
  \left[D^2+\frac{1}{r^2}(|\chi|^2-1) + \frac{1}{2}\,  |\phi |^2 ~ \right]\chi=
  -\frac{i}{2}\,  \, \phi^2
  \label{chiEq}
  \end{equation}
  \begin{equation}
  \left[D^\mu r^2 D_\mu+\frac{1}{2}(|\chi|^2+1) +
  2\lambda r^2 \left(|\phi|^2- 1 \right)
  \right] \phi= i \, \chi \phi^* \ .
  \label{phiEq}
  \end{equation}
  \label{gvarEqs}
\end{mathletters} \hskip -5pt
}
\noindent
To solve these equations given an initial configuration, we
must specify the appropriate boundary conditions. Boundary
conditions for the fields at $r=0$ can be derived from the
requirement that the \hbox{4-dimensional} configurations
they parameterize be regular at the origin. One finds that the
behavior as $r \to 0$ must be
\begin{mathletters}
\label{zeror}
\begin{eqnarray}
  a_0 &=& a_{0,1} r +  a_{0,3} r^3 +\dots \label{zerora}\\
  a_1&=& a_{1,0} + a_{1,2} r^2 +a_{1,4} r^4 + \dots \label{zerorb}\\
  \alpha&=& \alpha_1 r + \alpha_3 r^3 +  \alpha_5 r^5 +\dots \\
  \beta&=&-1 + \beta_2 r^2 + \dots \\
  \mu&=&\mu_0 + \mu_2 r^2 + \dots \\
  \nu&=&\nu_1 r + \nu_3 r^3 + \dots \ ,
\end{eqnarray}
\end{mathletters} \hskip -3pt
where the coefficients of the $r$-expansion are undetermined
functions of time. The $r$ behavior of the various fields is
determined by the requirement that $r=(x^2+y^2+z^2)^{1/2}$
have the appropriate power to render \hbox{4-dimensional}
fields analytic in $x$, $y$ and $z$. For example, since $A_0$
is proportional to $a_0 \lpmb{\sigma}{\bf \cdot \hat x}=
(a_0/r)\lpmb{\sigma}{\bf \cdot x}$, $a_0$ must be odd
in $r$. In terms of $\chi$ and $\phi$, the boundary conditions
at $r=0$ become
\begin{mathletters}
\label{bczeror}
\begin{eqnarray}
 a_0(0,t) &=& 0 \label{abc}\\
 \chi(0,t)&=&- i  \label{chibc}\\
 {\rm Re}\, \partial_r\phi(0,t)&=&0  \label{rephibc}\\
 {\rm Im} \, \phi(0,t)&=& 0 \label{imphibc} \ .
\end{eqnarray}
\end{mathletters} \hskip -4pt
Since $\Omega(r,t)$ vanishes at the origin,
one can check that these
boundary conditions are gauge invariant under
spherical gauge transformations.

There is an additional $r=0$ boundary condition given by
\begin{equation} \label{rzerobc}\\
a_{1,0}=\alpha_1 \ ,
\end{equation}
which is obtained by requiring that the two terms in
(\ref{sphai}) proportional to $\lpmb\sigma \cdot {\bf \hat x}$
cancel as $r\to 0$. Note that the $\nu=0$ component of
(\ref{fEq}) is the Gauss' law constraint, and once imposed
on the initial data it remains satisfied at subsequent times.
Substituting (\ref{zeror}) into Gauss' law gives
\hbox{$\partial_t (a_{1,0}-\alpha_1)=0$}. Therefore, if the
boundary condition $a_{1,0}=\alpha_1$ is satisfied by the
initial data it remains satisfied.

We turn now to large-$r$ boundary conditions. Since we
are interested in finite energy solutions, we require
that the fields go to pure gauge at large $r$. Hence, from
(\ref{sphvac}), \hbox{$a_\mu \to \partial_\mu\Omega$},
\hbox{$\chi \to -i \, \exp[i \Omega]$} and
\hbox{$\phi \to \pm \, \exp[i \Omega/2]$} as $r\to\infty$,
where $\Omega(r,t)$ is the spherical gauge function
defined in (\ref{Usph}). We can choose a gauge in
which $\Omega$ at spatial infinity becomes a constant,
independent of $r$ and $t$, so that $a_\mu \to 0$ as $r\to\infty$.
When we compactify \hbox{3-space} and require $\Omega \to 2 n
\pi$ at large $r$ for integer $n$, then $\chi \to -i$ and
$\phi \to \pm 1$ as $r \to \infty$. But as discussed
in the previous section this is inconvenient for parameterizing
the sphaleron, and instead we will take $\Omega \to (2 n+1)\pi$
for integer $n$. Then the \hbox{4-dimensional} gauge function $U$
maps spatial infinity onto the equatorial sphere of $SU(2)$, and
we cannot compactify space. In this case, however, $\chi \to i$
and $\phi \to \pm \, i$ as $r\to\infty$. We will choose the
plus sign for $\phi$, and in summary we take the large-$r$
boundary conditions to be
\begin{mathletters}
\label{bclarger}
\begin{eqnarray}
  a_\mu(r,t) & \to & 0 \label{alarger}\\
  \chi(r,t)  &\to & i  \label{chilarger}\\
  \phi(r,t) &\to& i  \label{philarger}
\end{eqnarray}
\end{mathletters} \hskip -5pt
as $r\to\infty$. There will be times in which it is convenient,
mostly for purposes of illustration, to take the boundary conditions
$a_\mu \to 0$, $\chi \to -i$ and $\phi \to 1$ as $r \to \infty$
consistent with spatial compactification, however, unless
otherwise specified, we will use the boundary conditions
(\ref{bclarger}).

One can now solve the equations of motion for initial
configurations and investigate to what extent topology
changing transitions occur. Since one cannot obtain
analytic solutions, we will exploit computational methods.
These numerical techniques, which are presented in the
next section, are based on a Hamiltonian formulation, so we
close this section with a brief exposition of the Hamiltonian
approach to the continuous system.

Central to this approach are the conjugate momenta to
the fields, defined by
\begin{mathletters}
\label{conjmom}
\begin{eqnarray}
  E &&\, \equiv \frac{1}{4\pi}\,\frac{\partial {\cal L}}{\partial
  \dot a_1} =   r^2(\partial_0 a_1 - \partial_1 a_0) \\
  \pi_\chi && \, \equiv \frac{1}{4\pi}\,\frac{\partial
  {\cal L}}{\partial \dot  \chi^*} = D_0 \chi \\
  \pi_\phi && \, \equiv \frac{1}{4\pi}\,\frac{\partial
  {\cal L}}{\partial \dot \phi^*} = r^2 D_0\phi \ ,
\end{eqnarray}
\end{mathletters} \hskip -4pt
where ${\cal L}$ is the Lagrangian density for the action
(\ref{effAction}). Since $a_0$ does not appear in
(\ref{effAction}), it has no corresponding conjugate
momentum and is not considered a dynamical variable.
Upon inverting (\ref{conjmom}) for the time derivatives
of the dynamical fields, the Hamiltonian of the system is
found to be $H+H_C$ where
  \begin{mathletters}
\label{HpHC}
  \begin{eqnarray}
  H =  4\pi \int^\infty_0dr \, && \bigg [  \,
  \frac{E^2}{2 r^2}   + |\pi_\chi|^2 +  |\pi_\phi|^2 +
  | D_r \chi |^2  + r^2 | D_r\phi |^2
  \nonumber\\
  && +\frac{1}{2 r^2}\left( ~ |\chi |^2-1\right)^2
  +\frac{1}{2}(|\chi|^2+1)|\phi|^2
  \nonumber \\
  &&  + {\rm Re}(i \chi^* \phi^2) +  \lambda r^2 \,
  \left(|\phi|^2 - 1\right)^2
  \bigg ]
  \label{SphHam}
  \end{eqnarray}
and
\begin{equation}
  H_C = 4\pi \int^\infty_0dr \
  a_0 \left[   -\partial_r E + i ( \pi_\chi^* \chi  -
  \chi^* \pi_\chi) +
  \frac{i}{2}\, ( \pi_\phi^* \phi  - \phi^* \pi_\phi) \right] \ .
\end{equation}
 \end{mathletters} \hskip -5pt

Variation with respect to $a_0$ gives Gauss' law
\begin{equation}
 \partial_r E = i ( \pi_\chi^* \chi  - \chi^* \pi_\chi) +
  \frac{i}{2}\, ( \pi_\phi^* \phi  - \phi^* \pi_\phi) \ \ .
\label{GaussLaw}
\end{equation}
Note that this is also the $\nu=0$ component of (\ref{fEq}). This
is not a dynamical equation just as $a_0$ is not a dynamical
variable. In fact, the Hamiltonian formulation makes it clear
that this equation is a constraint equation and $a_0$ is the
corresponding Lagrange multiplier. If the initial data are
chosen to satisfy Gauss' law, it will continue to be satisfied
at subsequent times.

In the $a_0=0$ gauge, the variables
\begin{equation}
a_1(r), \quad \chi(r), \quad \phi(r)
\label{achiphi}
\end{equation}
form a set of canonical coordinates conjugate to the momenta
\begin{eqnarray}
E(r)= &&r^2 \partial_0 a_1  \nonumber \\
\pi_{\chi}(r)=&&\partial_0 \chi  \nonumber \\
\pi_{\phi}(r)=&& r^2\partial_0 \phi \ .
\label{Err}
\end{eqnarray}
The evolution of these variables is generated by the Hamiltonian
(\ref{SphHam}). Gauss' law, (\ref{GaussLaw}), expresses the
residual invariance of the system under time independent local
gauge transformations and is imposed as a constraint on the
initial configuration.  It is subsequently conserved by the
equations of motion. Given initial data also satisfying the
regularity boundary condition $a_{1,0}=\alpha_1$, and using
the boundary conditions (\ref{bczeror}) and (\ref{bclarger}), a
regular solution is uniquely determined. We now turn to
approximating this solution numerically.

\section{Classical Evolution on the Lattice}
\hspace{1cm}
To solve the equations of motion numerically the system must
be discretized. For this purpose we subdivide the $r$-axis into
$N$ equal subintervals of length $\Delta r$ with finite length
$L= N \, \Delta r$. Thus, the lattice sites have spatial coordinates
$r_i=i\Delta r$ with $i=0 \cdots N$ (for our numerical simulations
we shall take $N=2239$ and $\Delta r = 0.04$, giving a lattice
of size $L=89.56$). It is convenient to use the formalism of lattice
gauge theories in assigning the space components of the gauge
fields to the oriented links between neighboring sites and in the
definition of gauge-covariant finite difference operators. For
simplicity, we will identify the lattice links via the midpoints
between lattice sites, which have coordinates
$r_{i+1/2}=(i+1/2)\Delta r$ with $i=0 \cdots N-1$.

The variables for the discretized system will now be defined
as follows. The zero component gauge degrees of freedom
are defined over the lattice sites, and are given by
\begin{equation}
  a_{0,i}(t) \quad {\rm for} \quad i=1 \dots N-1
\end{equation}
with $a_{0,0}=a_{0,N}=0$. The spatial components of the
gauge field are defined over the links of the lattice.
We will use the notation $a_{1,i}$, or simply $a_i$, to
represent the gauge variable defined over the link between
$r_i$ and $r_{i+1}$. This gives the variables
\begin{equation}
  a_i(t) \equiv a_{1,i}(t) \quad {\rm for}
  \quad i=0 \dots N-1 \ .
\end{equation}
As we show momentarily, boundary conditions for the spatial
variables $a_i$ are not required to determine the evolution of
the system. However, just as in the continuum, we will impose
an initial data boundary condition on $a_0$ corresponding to
(\ref{rzerobc}) to ensure the regularity of the four dimensional
fields at the origin (this condition will be discussed shortly).

The other field variables become
\begin{equation}
 \chi_i(t) \quad {\rm for} \quad i=1 \dots N-1 \ .
\end{equation}
with $ \chi_0=-i $, $\chi_N=i $
\medskip
and
\begin{equation}
 \phi_i(t) \quad {\rm for} \quad i=1 \dots N-1
\end{equation}
with $ \phi_N=i$. We are using boundary conditions at $r=L$
motivated by (\ref{bclarger}). These boundary conditions do
not admit spatial compactification and are chosen so that
perturbations about the sphaleron may be parameterized more
conveniently. Occasionally we will take the boundary conditions
$\chi_N = -i$ and $\phi_N = 1$ consistent with spatial
compactification; however, unless otherwise specified we will
use the aforementioned large-$r$ boundary conditions.

The value of $\phi$ at $r=0$ has so far not been specified.
We will return to this in a moment, but first we consider
the discretized covariant derivative. The time-like covariant
derivatives need no modification, but the continuum covariant
spatial derivatives are replaced by covariant finite
differences, e.g.
\begin{equation}
D_r \chi \to { \exp[- i a_i\, \Delta r] \,
\chi_{i+1}- \chi_i \over \ \Delta r}
{}~~~~~~~ i=0 \cdots N-1\ ,
\label{Drdisc}
\end{equation}
and like the gauge fields they are to be thought of as being
defined on the links between lattice sites. The rest of the
discretization is straightforward, and one obtains a discretized
action $S_D$ expressed in terms of a finite set of variables
which still possess an exact local gauge invariance:
\begin{mathletters}
\begin{eqnarray}
 a_{0,i} &\to& a_{0,i} + \partial_t \Omega_i  \hskip1.1in  i=0 \dots N
\\
 a_i &\to& a_i + (\Omega_{i+1}-\Omega_{i})/\Delta r
\hskip0.4in i=0 \dots N-1
\\
  \chi_i &\to& e^{i  \Omega_i} \chi_i \hskip1.45in  i=0 \dots N
\\
  \phi_i &\to& e^{i  \Omega_i/2} \phi_i \hskip1.33in   i=0 \dots N \ .
\end{eqnarray}
\end{mathletters} \hskip -5pt
The discretized gauge function $\Omega_i(t)$ with $i=0 \dots N$
is defined over the lattice sites, and satisfies $\Omega_0(t)=0$
to maintain the regularity of the corresponding
\hbox{4-dimensional} gauge transformed fields.

Before we continue, however, we must derive the boundary
condition for $\phi$ at $i=0$. This is obtained from (\ref{rephibc})
and (\ref{imphibc}), in which the continuum field $\phi$ at $r=0$
is real with vanishing spatial derivative. Since a statement
about the ``derivative'' is not gauge covariant, we prefer to
state that the real part of the covariant derivative $\partial_r
\phi - i a \phi$, together with the imaginary part of $\phi$,
must vanish at $r=0$. This is equivalent to (\ref{rephibc}) and
(\ref{imphibc}) since $\phi$ is real at $r=0$. But it has the
advantage that it translates into the following boundary
conditions for the discretized case:
\begin{mathletters}
\begin{eqnarray}
 {\rm Re} [\exp(\frac{-i \, a_0 \, \Delta r}{2}) \phi_1 - \phi_0 \ ]
  &=&0 \\
 {\rm Im} \phi_0 &=& 0 \ ,
\end{eqnarray}
\end{mathletters} \hskip -5pt
where $a_0$ is the value of $a_{1,i}$ at $i=0$ and should
not be confused with the time-like vector field. Thus, we
write the boundary condition as
\begin{equation}
  \phi_0 =  {\rm Re} [ \exp({-i \,a_0\, \Delta r \over 2}) \phi_1] \ ,
\label{phi0bc}
\end{equation}
which allows us to eliminate $\phi_0$ from the list of
dynamical variables.

The discretized Lagrangian becomes
\begin{eqnarray}
  L &=&4\pi \, \sum_{i=0}^{N-1} \bigg\{\frac{r^2_{i+1/2} }{2} \,
 \bigg(\partial_0 a_i   -\frac{a_{0,i+1}-a_{0,i}}{\Delta r}
 \bigg)^2 - \frac{ |\exp(-i\,a_i\, \Delta r)
  \chi_{i+1} -\chi_i|^2}{\Delta r^2} \bigg\} \Delta r
 \nonumber \\ &&
 + \,4\pi \, \sum_{i=1}^{N-1} \bigg\{ |(\partial_0-i a_{0,i})\chi_i|^2
 +r^2_i \, |(\partial_0-{i a_{0,i} \over 2})\phi_i|^2
 - r^2_{i+1/2}  \, \frac{|\exp({-i\,a_i\, \Delta r / 2})
 \phi_{i+1} -\phi_i|^2}{\Delta r^2 }
\nonumber \\ &&
  -\frac{1}{2} \, (|\chi_i|^2+1)|\phi_i|^2 -  {\rm Re}(i\chi_i^*
  \phi_i^2) - {1 \over 2 r^2_i}(|\chi_i|^2-1)^2  - \lambda
  r^2_i \, (|\phi_i|^2-1)^2  \bigg\} \Delta r
\nonumber \\ &&
- \,4\pi \, r^2_{1/2}{[{\rm Im}(\exp(-i\,a_0\, \Delta r / 2) \phi_1)]^2
 \over  \Delta r} \ .
\label{discL}
\end{eqnarray}
This Lagrangian was obtained by discretizing the system
as previously explained and by replacing $\phi_0$ by the
right hand side of (\ref{phi0bc}). One might think this induces
an additional contribution to the kinetic term of $\phi_1$ from
the time derivative of (\ref{phi0bc}). However, the term
proportional to $\dot \phi_0$ vanishes since it is multiplied
by $r^2_0=0$, and hence (\ref{discL}) is the complete
Lagrangian.

We define conjugate momenta (the factor $1/4\pi \Delta r$ is
introduced so as to have Poisson brackets with a continuum
like normalization $\lbrace \pi_i^*, \phi_j \rbrace =  \delta_{i,j}/
\Delta r $ etc.)

\begin{mathletters}
\label{congMom}
\begin{eqnarray}
  E_i &=& {1 \over 4\pi \Delta r}{ \partial L \over \partial (\partial_0 a_i)}
=
  r^2_{i+1/2} \,  (\partial_0 a_i -{a_{0,i+1}-a_{0,i} \over \Delta r})
\hskip0.2in  i=0 \dots N-1
\label{congMoma} \\
   P_i &=& {1 \over 4\pi \Delta r}{ \partial L \over \partial
(\partial_0 a_{0,i})} = 0 \hskip1.89in  i=0 \dots N
\label{congMomb} \\
  p_i &=& {1 \over 4\pi \Delta r}{ \partial L \over \partial (\partial_0
\chi_i^*)} =
  \partial_0 \chi_i -i a_{0,i} \chi_i  \hskip1.09in  i=0 \dots N
\label{congMomc} \\
  \pi_i &=& {1 \over 4\pi \Delta r}{ \partial L \over \partial
 (\partial_0 \phi_i^*)} = r^2_i(\partial_0 \phi_i -{ i a_{0,i} \over 2}
 \phi_i )  \hskip0.81in  i=0 \dots N \ .
\label{congMomd}
\end{eqnarray}
\end{mathletters} \hskip -5pt
Equation (\ref{congMomb}) is a primary constraint equation, in
the sense of Dirac. {}From (\ref{discL}) and (\ref{congMom}) we
obtain the Hamiltonian  $H+H_C$, with
\begin{mathletters}
\label{discH}
\begin{eqnarray}
  H&=&4\pi \,\sum_{i=0}^{N-1} \bigg\{ { E_i^2 \over 2 r^2_{i+1/2}} +
  {|\exp(-i\,a_i\, \Delta r) \chi_{i+1} -\chi_i|^2 \over \Delta r^2 }
  \bigg\} \, \Delta r
  + \,4\pi\,\sum_{i=1}^{N-1} \bigg\{  |p_i|^2  + { |\pi_i|^2 \over r^2_i}
\nonumber  \\ &&
  + r^2_{i+1/2}{|\exp(-i\,a_i\,\Delta r/ 2) \phi_{i+1} -\phi_i|^2
  \over \Delta r^2} +{1 \over 2} (|\chi_i|^2+1)|\phi_i|^2+
  {\rm Re}(i\chi_i^*\phi_i^2)
 \nonumber \\ &&
  +{1 \over 2 r^2_i} (|\chi_i|^2-1)^2
  +\lambda \,  r^2_i(|\phi_i|^2-1)^2 \bigg\} \, \Delta r
  + \,4\pi\,r^2_{1/2}{[{\rm Im}(\exp(-i\,a_0\, \Delta r/2)
  \phi_1)]^2 \over \Delta r}
\label{discHa}
\end{eqnarray}
and
\begin{eqnarray}
  H_C= 4\pi \, \sum_{i=1}^{N-1} a_{0,i}
  \bigg\{ - \frac{E_i - E_{i-1}}{\Delta r}
  +i (p_i^*\chi_i -\chi_i^* p_i)
  +{i \over 2} (\pi_i^*\phi_i -\phi_i^*\pi_i) \bigg\} \,\Delta r \ .
\label{discHb}
\end{eqnarray}
\end{mathletters} \hskip -5pt
Upon commuting (or more precisely, taking the Poisson bracket)
the constraint (\ref{congMomb}) with $H+H_C$ one obtains as
a further constraint Gauss' law
\begin{equation}
  {E_i - E_{i-1}\over \Delta r} =
   i (p_i^*\chi_i -\chi_i^* p_i)
  +{i \over 2} (\pi_i^*\phi_i -\phi_i^*\pi_i)
  \equiv j_i \ , \quad i=1 \dots N-1 \ .
\label{discGL}
\end{equation}
We impose the second class constraint
$a_{0,i}=0$ for $i=1 \dots N-1 $.
The equations of evolution that follow from H are then
\begin{mathletters}
\label{NLMomEq}
\begin{eqnarray}
{da_i \over dt} &= &{ E_i \over r^2_{i+1/2} } \hskip0.8in  i=0 \dots N-1 \\
{d\chi_i \over dt} &=&  p_i   \hskip1.1in  i=1 \dots N-1 \\
 {d\phi_i \over dt} &=& { \pi_i \over r^2_i }, \hskip0.975in  i=1 \dots N-1
\end{eqnarray}
\end{mathletters} \hskip -5pt
and

\begin{mathletters}
\label{NLVarEq}
\begin{eqnarray}
  {dE_i \over dt} &=& i {\chi_{i+1}^* \exp(i a_i\, \Delta r)
  \chi_i - \chi_i^* \exp(- i a_i\, \Delta r) \chi_{i+1} \over \Delta r}
  \nonumber \\ &&
  + \, i {r^2_{i+1/2}\over 2} {\phi_{i+1}^* \exp(i a_i\,  \Delta r /2)
  \phi_i - \phi_i^* \exp(- i a_i\,  \Delta r /2) \phi_{i+1} \over  \Delta r}
  \quad i=0 \dots N-1
  \label {NLVarEqE}  \\
  {dp_i \over dt} &=&  {\exp(-i a_i\,  \Delta r) \chi_{i+1}
  - \chi_i \over  \Delta r^2} +
  {\exp(i a_{i-1}\,  \Delta r) \chi_{i-1} - \chi_i \over  \Delta r^2}
  \nonumber \\ &&
  - {\chi_i |\phi_i|^2 +i \phi_i^2 \over 2}
  - \frac{1}{r^2_i}\chi_i (|\chi_i|^2 - 1)
  \hskip1.77in  i=1 \dots N-1
 \label {NLVarEqpi}  \\
  {d\pi_i \over dt} &=& r^2_{i+1/2} {\exp(-i a_i\,  \Delta r / 2)
  \phi_{i+1} - \phi_i \over  \Delta r^2} + r^2_{i-1/2}
  {\exp(i a_{i-1}\,  \Delta r / 2) \phi_{i-1} - \phi_i \over  \Delta r^2}
  \nonumber \\ &&
  - {\phi_i (|\chi_i|^2+1)\over 2} + i \chi_i \phi_i^*
  -2 \lambda \, r^2_i \phi_i (|\phi_i|^2 - 1)
  \hskip1.05in i=1 \dots N-1 \ ,
\label {NLVarEqphi}
\end{eqnarray}
\end{mathletters} \hskip -3pt
where $\phi_0$ is given by (\ref{phi0bc}), $\phi_N=i$, $\chi_0=-i$
and $\chi_N=i$ (or $\chi_N=-i$ and $\phi_N=1$, if as we will
occasionally do, boundary conditions consistent with spatial
compactification are used). The momenta of $\chi$ and $\phi$
vanish at $i=0$ and  $i=N$.

In summary, we have the following table of independent dynamical
variables and their respective conjugate momenta:

\bigskip
\centerline{
  \begin{tabular}{|c|c|c|c|}
  \hline
  $\hbox{~variable~}  $ &  $\hbox{~momentum~} $ &
 $\hbox{~index range~} $
  & $ \hbox{number}  $  \\
  \hline
 $a_i$ & $E_i$ & $~i=0 \dots N-1~$ & $N$  \\
  \hline
  $\chi_i$ & $p_i$ & $~i=1 \dots N-1~$ & $~2(N-1)~$  \\
  \hline
  $\phi_i$ & $\pi_i$ & $~i=1 \dots N-1~$ & $~2(N-1)~$  \\
  \hline
\end{tabular}
}
\bigskip

\noindent
Since we have set $a_{0,i}$ to zero, the number of dynamical
variables and momenta (excluding boundary fields at $r=0$ and
$r=L$) are $2(5N-4)$. Note that (\ref{NLMomEq}) and  (\ref{NLVarEq})
give $2(5N-4)$ equations, so the system is uniquely determined
given the initial values of the fields $\chi$ and $\phi$ and their
momenta (note that boundary conditions for the spatial gauge field
$a_i$ are not required). The initial data must be chosen to be
consistent with Gauss' law (\ref{discGL}). We will also impose the
boundary condition $a_0={\rm Re} (\chi_1-\chi_0) / \Delta r $, which
approximates the continuum relation (\ref{rzerobc}). (This relation,
which would be conserved in the continuum limit, will remain
satisfied to $O(\Delta r)$ in the evolution of the discretized system).

The restriction to uniform spacing of the subintervals on the $r$-axis
is not fundamental and we have also implemented a discretization in
which $\Delta r$ increases as one moves out on the $r$-axis. In this
manner one can effectively make the system larger and delay the
effects of the impact of the waves with the boundary without
worsening the spatial resolution near $r=0$, where most of the
non-linear dynamics takes place. We have found, however, that the
advantages one gains hardly warrant the additional complications
introduced by the non-uniform spacing.

For the numerical integration of the time evolution we have used the
leap-frog algorithm. Since this algorithm constitutes one of the
fundamental techniques for the integration of ordinary differential
equations of the Hamiltonian type and as such is textbook material,
we will not discuss it in depth. Essentially, given conjugate
canonical variables $q_i$ and $p_i$ which obey equations
\begin{eqnarray}
{dq_i \over dt} &=& g_i(p)
\nonumber\\
{dp_i \over dt} &=& f_i(q) \ ,
\label{masteq}
\end{eqnarray}
one evolves the values of $q$ and $p$ from some initial $t$ to
$t+\Delta t$ as follows. In a first step $p_i$ is evolved to the
mid-point of the time interval by
\begin{eqnarray}
p_i \to p_i'=&& p_i+f_i(q) {\Delta t \over 2}
\nonumber \\
q_i \to q_i'=&& q_i
\label{lfone}
\end{eqnarray}
(although $q_i$ is left unchanged, it is convenient to consider the
step formally as a transformation of the entire set of canonical
variables). In a second step one evolves the coordinates from
their initial value $q_i=q_i'$ to their value at the end of the interval
\begin{eqnarray}
p_i' \to p_i''=&& p_i' \nonumber \\
q_i' \to q_i''=&& q_i' +g_i(p') \Delta t \ .
\label{lftwo}
\end{eqnarray}
Finally, the momenta are evolved from their value at the midpoint
to the final value
\begin{eqnarray}
p_i'' \to p_i'''=&& p_i''+f_i(q'') {\Delta t \over 2} \nonumber \\
q_i'' \to q_i'''=&& q_i'' \ .
\label{lfthree}
\end{eqnarray}

One can easily verify that these equations reproduce the correct
continuum evolution from $t$ to $t+\Delta t$ up to errors of order
$(\Delta t)^3$. Moreover, the algorithm has the very nice property
that all three steps above constitute a canonical transformation
and that it is reversible (in the sense that starting from $q_i'''$,
$-p_i'''$, up to round-off errors one would end up exactly with
$q_i$, $-p_i$). Because the physical solutions of interest are the
time reversed processes of the ones we numerically evolve, it is
important that we use an algorithm that is reversible. Another
very nice feature of the algorithm is that, although the
evolution of the variables is affected by errors of order
$(\Delta t)^3$, the energy of a harmonic oscillator, and
therefore of any system which can be decomposed into a linear
superposition of harmonic oscillators, is conserved exactly
(always up to round-off errors, but if one works as we do in
double precision, these are very small). Since extracting the
asymptotic normal mode amplitudes is the heart of our numerical
approach, it is also important to have an algorithm that is well
behaved in the linear regime. One final comment is in order. In a
sequence of several iterations of the algorithm, after the
momenta have been evolved by the initial $\Delta t/2$, the first and
third steps, (\ref{lfone}) and (\ref{lfthree}) respectively, can be
combined into a single step, whereby the momenta are evolved from
the midpoint of one interval to the midpoint of the next one ``hopping
over'' the coordinates, which are evolved from endpoint to endpoint.
This motivates the name assigned to the algorithm.

\section{The Initial Configuration: Perturbation about
the Sphaleron }
\hspace{1cm}
With a good grasp on numerical solutions of the equations of
motion, we can turn now to the second crucial component of
the computation, namely the parameterization of the initial
configuration. One could easily construct an initial state
consisting of an incoming wave in the linear regime; however,
it would be very difficult to ensure that such a configuration
underwent a topology change during its subsequent evolution.
Instead, it is much more convenient to parameterize the initial
state at or near the instant of topology change. The system is
then allowed to evolve until the linear regime is reached, at
which point the particle number can be extracted in the manner
explained in the next section. The physical process of interest
is then the time reversed solution, which starts in the
linear regime with a known particle number and undergoes
a change of topology at subsequent times. (In fact, it must
be explicitly checked that the winding number of the outgoing
configuration is different from the incoming one, ensuring that
the topology has changed, since the system could pass back
over the sphaleron barrier and into the original
topological sector. We have found however that topology
change does typically occur.)

Topology changing transitions within the spherical {\it Ansatz}
are characterized by the vanishing of $\phi$ at $r=0$ and
the vanishing of $\chi$ at nonzero $r$. The zero of $\chi$ is
reminiscent of the zero which characterizes the sphaleron of the
Abelian Higgs model. However, as shown in Ref.~\cite{fggrs},
it is the zero of the Higgs field (i.e.~the zero
of $\phi$) which carries a deeper significance and should be
associated with the actual occurrence of the topological
transition. For a sequence of configurations that pass directly
through the sphaleron these two zeros occur at the same time.
Nonetheless, this is not the most general case and the zeros of
$\phi$ and $\chi$ need not occur simultaneously (although for
a topological transition, {\it both} fields will vanish sometime
during their evolution)\cite{gi}. We are free then to parameterize
initial topology changing configurations imposing that either
$\phi$ vanish at the origin or that $\chi$ has a zero at some
non-zero $r$. It is convenient to choose the latter, in which we
parameterize the initial configuration in terms of coefficients $c_n$
of some suitable expansion of the fields and their conjugate
momenta, constrained only by the boundary conditions and the
requirement that the field $\chi$ has a zero at some non-zero $r$.
Furthermore, we can use the residual time independent gauge
invariance to make $\chi$ pure imaginary at the initial time. The
field $\phi$ is only restricted to obey the boundary conditions and
does not necessarily vanish at the origin (although it will vanish
at the origin at some instant in its evolution if the topology is
to change).

To be more specific, we parameterize each field as a (not necessarily
small) perturbation about the sphaleron given by a linear combination
of spherical Bessel functions with the appropriate small-$r$ behavior
of (\ref{zeror}). We only need the first three functions,
\begin{mathletters}
\label{sphbess}
\begin{eqnarray}
j_0(x) &=& \frac{\sin x}{x}  \label{jzero} \\
j_1(x) &=& \frac{\sin x}{x^2} - \frac{\cos x}{x}  \label{jone} \\
j_2(x) &=&   \left(\frac{3}{x^3} - \frac{1}{x}\right) \sin x -
\frac{3}{x^2} \cos x \label{jtwo} \ ,
\end{eqnarray}
\end{mathletters} \hskip -3pt
since $j_0(x)\sim 1$, $j_1(x)\sim x$ and $j_2(x)\sim x^2$ at
small $x$. Motivated by the boundary conditions
(\ref{bclarger}), we require the perturbation to vanish at $r=L$.
We thus parameterize perturbations about the sphaleron
in terms of \hbox{$j_{n m}(r)=j_n(\alpha_{n m}r/L)$} with $n=0,1$
or $2$, where $\alpha_{n m}$ are the zeros of $j_n(x)$, i.e.
$j_n(\alpha_{n m})=0$ with $m=1,2,\cdots$.
The functions $j_{n m}(r)$ form a complete set for every
$n$, and the small-$r$ behavior determines
the appropriate value of $n$ for each field.
The reader should note that the expansion of the
starting configuration in terms of Bessel functions is largely
a matter of convenience. This expansion is not related to the
expansion of the fields in the linear regime (to be discussed
in the next section), and any complete set of
functions with the correct behavior as $r\to 0$ can be used to
parameterize a perturbation of the sphaleron localized in the
neighborhood of the origin.

Recall that we must impose the boundary condition $a_{1,0}=
\alpha_1$ on the initial data (using continuum notation). We
are working in the $a_0=0$ gauge, but we still have the freedom
to impose a time independent gauge transformation on the starting
configuration to set $\alpha=0$. Therefore, (\ref{zerorb}) gives
\hbox{$a_1(r)\sim r^2$} at small $r$, and hence $a_1(r)$ is
expanded only in terms of $j_2(x)$. We are thus led to parameterize
the initial configuration by

\vbox{
\begin{mathletters}
\label{fieldparam}
\begin{eqnarray}
\chi(r) &=& \chi_{\rm sph}(r) + i \sum_{m=1}^{N_{\rm sph} } \, c_{1 m}\
j_{2 m}(r) \label{chiparam} \\
\phi(r) &=& \phi_{\rm sph}(r) + \sum_{m=1}^{N_{\rm sph} } \, c_{2 m}\
 j_{0 m}(r)+ i \sum_{m=1}^{N_{\rm sph} } \, c_{3 m}\  j_{1 m}(r)
\label{phiparam} \\
\pi_\chi(r) &=& \sum_{m=1}^{N_{\rm sph} } \, c_{4 m}\
 j_{1 m}(r)+ i \sum_{m=1}^{N_{\rm sph} } \, c_{5 m}\  j_{2 m}(r)
\label{pichiparam} \\
\pi_\phi(r) &=& \left[ \, \sum_{m=1}^{N_{\rm sph} }\, c_{6 m}\
 j_{0 m}(r)+ i \sum_{m=1}^{N_{\rm sph} } \, c_{7 m}\  j_{1 m}(r)  \,
\right ] \, r^2
\label{piphiparam} \\
a_1(r) &=& \sum_{m=1}^{N_{\rm sph} } \,c_{8 m}\ j_{2 m}(r) \ ,
\label{aparam}
\end{eqnarray}
\end{mathletters} \hskip -5pt
}
\noindent
where $\chi_{\rm sph}=i(2 f -1)$ and $\phi_{\rm sph}=i h$ as in
(\ref{sphSphal}), and where we have cut off the sums
at some $N_{\rm sph} \le N$. The most general initial
configuration is obtained with $N_{\rm sph} = N$, but to avoid
exciting short wavelength modes which only correspond to
lattice artifacts, we take $N_{\rm sph} < N/5$ to $N/10$.
This implies no limitations on the physical properties of the
system other that those coming from an ultraviolet
cutoff (finite $\Delta r$) anyway, and as one expects
this is born out by numerical results in which typical
solutions excite only modes with wavelength substantially
larger than the lattice spacing. As the dimension of the
initial configuration space is $8N_{\rm sph}$, and since
the lattice we work with is rather large, to improve the
efficiency of our stochastic search we have taken $N_{\rm sph}
\sim N/50$ ($N_{\rm sph}= 50$ for $N=2239$).

To obtain the correct small-$r$ behavior of $\pi_\phi$, we have
inserted an explicit factor of $r^2$ in (\ref{piphiparam}) because
$\pi_\phi = r^2 \partial_0 \phi$. The profile functions
$f$ and $h$ satisfy the boundary conditions $f(0)=h(0)=0$ and
$f(L)=h(L)=1$, and will be specified momentarily. For now it is
sufficient to note that since $\chi(0)=-i$ and $\chi(L)=i$,
and since $\chi(r)$ is pure imaginary, it will necessarily have
a zero for some $r>0$. Hence, (\ref{fieldparam}) specifies a
configuration at the moment in which $\chi$ vanishes. We should
also point out that because of its large-$r$ behavior,
(\ref{fieldparam}) is expressed in a gauge in that is
inconsistent with spatial compactification.

We have so far used continuum notation, but (\ref{fieldparam})
is to be understood as determining the configuration at the lattice
sites $r=r_i$ for \hbox{(\ref{chiparam})-(\ref{piphiparam})} and
at $r=r_{i+1/2}$ for (\ref{aparam}), i.e.~$\chi_i=\chi(r_i)$,
$p_i=\pi_\chi(r_i)$, $\phi_i=\phi(r_i)$, $\pi_i=\pi_\phi(r_i)$
and $a_i=a_1(r_{i+1/2})$. We have not yet specified the
electric field, but since the initial configuration must satisfy
Gauss' law we can determine $E_i$ by integrating
 (\ref{discGL}) outward from $i=0$ to $i=N-1$. The value of
$E_0$ must be given for this procedure however. In the
continuum $E(r=0)=0$, so one is tempted to set $E_0=0$.
But since $E_0$ lives on the first link at $r=r_{1/2}=\Delta r/2$,
it is better to set
\begin{eqnarray}
E_0 = \frac{\Delta r}{2} \, \frac{j_0+j_1}{2} =
 \frac{\Delta r \left[ i (p_1^*\chi_1 -\chi_1^* p_1)
 +i (\pi_1^*\phi_1 -\phi_1^*\pi_1)/2 \right] }{4} \ ,
\end{eqnarray}
and then subsequent values of $E_i$ for $i=1 \cdots N-1$
can be obtained by integrating (\ref{discGL}).

The sphaleron $\chi_{\rm sph}$, $\phi_{\rm sph}$ of (\ref{sphSphal})
is parameterized by profile functions $f(r)$ and $h(r)$ and is
a saddle point of the potential energy functional with one
unstable direction. This direction involves an excitation of the
\hbox{2-dimensional} gauge potential $a_1$. Hence the sphaleron
is an absolute minimum of the potential obtained from
(\ref{discHa}) by dropping the $a_1$ terms (and all the momenta).
Using the method of conjugate gradients, with an initial
guess for $f$ and $h$ that satisfies the appropriate boundary
conditions, we can obtain an extremely accurate approximation
to the sphaleron by minimizing
\begin{mathletters}
\begin{eqnarray}
 H_{\rm sph}/4\pi &= &\sum_{i=0}^{N-1} \bigg\{
 { |\chi_{i+1} -\chi_i|^2 \over \Delta r^2 }
 +r^2_{i+1/2}{ |\phi_{i+1} -\phi_i|^2
 \over \Delta r^2} +{1 \over 2} (|\chi_i|^2+1)|\phi_i|^2+
\nonumber \\
&& \qquad {\rm Re}(i\chi_i^*\phi_i^2)
+\lambda \, r^2_i(|\phi_i|^2-1)^2 \bigg\} \, \Delta r
+ \sum_{i=1}^{N-1} {1 \over 2 r^2_i} (|\chi_i|^2-1)^2 \, \Delta r
\\
&= &\sum_{i=0}^{N-1} \bigg\{  { 4(f_{i+1} -f_i)^2 \over \Delta r^2 }
 +r^2_{i+1/2}{ (h_{i+1} -h_i)^2
 \over \Delta r^2} + 2 (f_i-1)^2h_i^2 +
\nonumber \\
&& \qquad\qquad +\lambda \, r^2_i(h_i^2-1)^2 \bigg\} \, \Delta r
+ \sum_{i=1}^{N-1}{8 \over r^2_i} f_i^2 (1-f_i)^2 \, \Delta r \ ,
\label{sphpot}
\end{eqnarray}
\end{mathletters} \hskip -4pt
where we have used the boundary condition $\phi_0=
{\rm Im}\phi_1$ to extend the sum on $\phi$ in (\ref{discHa})
to include $i=0$. In our units and with $g=1$, the energy of
the sphaleron is then given by $\epsilon_{\rm sph} / 4\pi =
2.5426$ for $\lambda=0.1$.

\vbox{
\begin{figure}
\centerline{
\epsfxsize=80mm
\epsfbox{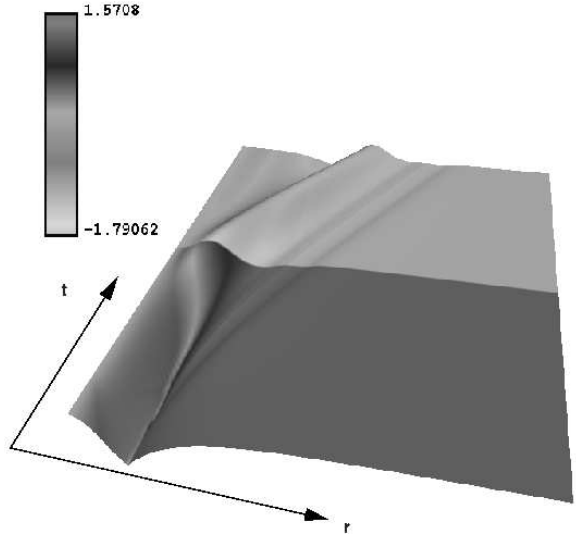}
}
\caption{\tenrm
Decay of a small perturbation about the sphaleron:
evolution of the $\chi$ field. The values of the phase
of the complex field are coded by different shades of gray,
and the modulus of the field by the height of the surface. As
explained in the next section, the asymptotic linear system
has a particle number of order $53$. The lattice parameters
are $N=2239$ and $\Delta r=0.04$ with a Higgs coupling
of $\lambda=0.1$. The initial configuration is given by
(5.2) with $c_{4,1}=0.00247$ being the only nonzero parameter.
For a full color figure see
http://cthulu.bu.edu/$\scriptstyle\sim$bobs/bviolate.html.
}
\end{figure}
}

We are now in a position to numerically evolve perturbations
about the sphaleron. Figure~5 illustrates the behavior of the
$\chi$ field for an initial configuration given by
(\ref{fieldparam}) with \hbox{$c_{4, m=1}=0.00247$}
and all other $c$-parameters zero. This is in fact the
configuration from which we have chosen to seed the
stochastic sampling procedure which we will describe
in Sec.~VII. We have found it very convenient and
informative to use color to code the phase of the complex
fields. Unfortunately the illustrations in these pages cannot
be reproduced in color and we have tried to render the
variation of the phase with a gray scale. At some point a
gauge transformation has been performed in Fig.~5
bringing the asymptotic linear state into the sector of
zero winding number (consistent with spatial compactification).
The gauge transformation is made manifest by the sudden
change of shading of the surface. We have performed
this gauge transformation because eventually we want to
study the topology change of the time reversed solution
(cf.~Fig.~6 below), and this is best done in a gauge in
which the asymptotic linear state has zero winding number.
Moreover, the gauge transformation also serves to give
a graphic illustration of the gauge invariance of our
procedure, which is made manifest by the fact that
although the shading (or color) of the surface changes,
there is no discontinuity in the surface itself.

{}From Fig.~5 it is clear that the energy, which is concentrated
in the neighborhood of $r=0$, disperses and gives rise to a
pattern of outgoing waves. The waves soon become linear
and possess a definite particle number, in this case of order
$53$ physical particles (using units appropriate to the
standard model, which we will refer to as physical units).

The physical process of interest is then the time reversed
solution which starts in the linear regime with known particle
number, proceeds through the non-linear sphaleron perturbation
(\ref{fieldparam}) at intermediate times and finally linearizes
once again at late times. Because of time invariance of the
equations of motion, this process can be obtained by first
evolving the perturbation (\ref{fieldparam}) until the linear
regime is reached, and then reversing the momenta and
evolving that configuration forward in time. The resulting
solution retraces the evolution of the sphaleron decay,
and then proceeds over the barrier into another topological
sector. Since our numerical strategy for obtaining asymptotically
linear topology changing solutions relies upon first evolving
the sphaleron perturbation, we shall refer to (\ref{fieldparam})
as the ``initial'' state, while the asymptotic linear states of the
physical process will be called the ``in'' and ``out'' states.

\vbox{
\begin{figure}
\centerline{
\epsfxsize=80mm
\epsfbox{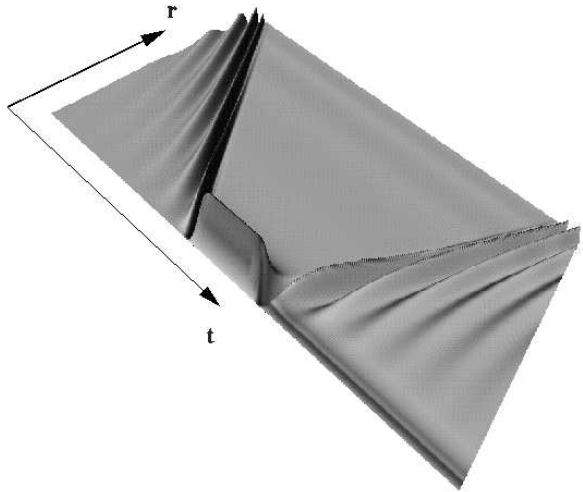}
}
\caption{\tenrm
Topology changing transition: behavior of the $\chi$ field
obtained from Fig.~5 by the time reversal procedure described
in the text. The various shades of gray code the phase of the
complex field. The field starts as an excitation about the trivial
vacuum, passes over the sphaleron and then emerges as an
excitation about the vacuum of unit winding. Note the persistent
strip of $2\pi$ phase change near $r=0$ after the wave bounces
off the origin. For a full color figure see
http://cthulu.bu.edu/$\scriptstyle\sim$bobs/bviolate.html.
}
\end{figure}
}

Figure~6 represents a physical process obtained from Fig.~5
in the above manner, and it illustrates the evolution of the
$\chi$ field for a topology changing solution. The ``initial''
state in Fig.~6, determined from (\ref{fieldparam}) by the
coefficients $c_n$, corresponds to the time-slice half way
through the depicted evolution. We have reverted to a gauge in
which the boundary conditions are $\chi_N=-i$ and $\phi_N=1$,
consistent with spatial compactification, and in which the
in-state has no winding and the out-state has unit winding
number. This process represents an imploding spherical energy
shell that converges on the origin, where a change of topology
takes place. The topology change is indicated by the strip of
rapidly varying tonality which persists in the neighborhood of
the origin and codes the variation of the $2\pi$ phase change of
$\chi$. With color, this strip would appear as a vivid rainbow,
left over as a marker of the change of topology of the evolving
fields.

It is important to keep in mind that an arbitrary configuration
(\ref{fieldparam}) does not necessarily produce a topology
changing solution, in the sense that at late times the out-state
might evolve back into the original topological sector. With our
parameterization (\ref{fieldparam}), however, we have found
that the system does in fact typically change topology.
Nonetheless, using the time reversed procedure above we can
always verify whether the in- and out-states have the same
topology, and if so the initial configuration that produced them
can be rejected (or equivalently, and more efficiently, we can
evolve the initial configuration (\ref{fieldparam}) both forward
and backward in time and compare the asymptotic states obtained
in this way).

We now have a procedure for constructing solutions which, in
the course of their evolution, undergo changes of topology. By
varying the values of the parameters $c_n$ we will be able to
study the properties of such field evolution and, in particular,
explore the domain of permissible values for $\epsilon$ and $\nu$.
Before we can implement this procedure, however, we must
devise a way to calculate the particle number in the asymptotic
linear regime. In the next section we describe how this can be
done.

\section{Normal Modes}
\hspace{1cm}
Given an initial configuration parameterized by the coefficients
$c_n$, we evolve the system until the linear regime is reached,
where the fields undergo small oscillations about a vacuum
configuration. The normal mode amplitudes $a_n$ may then be
extracted and the particle number computed using (\ref{nsum}).
We turn now to the problem of identifying the normal modes.

Since we have put the field theoretic system of interest on a
spatial lattice, to be entirely consistent we should also solve
the normal modes problem on the lattice. The discrete problem,
however, cannot be solved analytically and one must resort to
numerical methods. On the other hand, the normal modes of the
continuum system, even restricted to a box of finite size
$L=N\Delta r$, can be found analytically. We have solved the
problem both numerically on the lattice and analytically in the
continuum limit. The lattice we consider ($N=2239$ with
$\Delta r=0.04$) is big enough that there is excellent numerical
agreement between the normal modes found by the two methods
(the difference between the normalized modes never exceeds
$10^{-6}$), so we will present here only the continuum solution.

Following Ref.~\cite{gi}, we work in terms of gauge invariant
variables. We write the fields $\chi$ and $\phi$ in polar form,
\begin{mathletters}
\begin{eqnarray}
\chi &=& -i \rho \, e^{i\theta} \\
\phi &=& \sigma \, e^{i\eta} \ .
\end{eqnarray}
\end{mathletters} \hskip -5pt
The variables $\rho$ and $\sigma$ are gauge invariant.
We can also define the gauge invariant angle
\begin{eqnarray}
\xi=\theta-2\eta.
\end{eqnarray}
Finally, in \hbox{1+1} dimensions we can write
\begin{eqnarray}
r^2 f_{\mu\nu} = - 2 \epsilon_{\mu\nu} \psi
\end{eqnarray}
where $\epsilon_{01}=+1$ and $\mu$, $\nu$ run over
$0$ and $1$. The variable $\psi$ is gauge invariant. Rather
than working with the six gauge-variant degrees of freedom
$\chi$, $\phi$ and $a_\mu$ we use the four
gauge invariant variables $\rho$, $\sigma$, $\psi$ and
$\xi$. These variables satisfy the equations
\cite{gi}
\vbox{
\begin{mathletters}
\label{gieq}
\begin{equation}
 \partial_\mu\partial^\mu\rho
 - \frac{\rho \, (\frac{1}{4} r^2\sigma^2\partial_\mu\xi
 +\epsilon_{\mu\nu}\partial^\nu\psi )^2}{(\rho^2 +
 \frac{1}{4}r^2\sigma^2)^2} + \frac{1}{r^2}( \rho^2-1)\rho
 +\frac{1}{2} \rho \, \sigma^2 - \frac{1}{2} \, \sigma^2\cos \xi = 0
\label{giRho}
\end{equation}
\begin{equation}
 \partial_\mu r^2\partial^\mu\sigma - \frac{\frac{1}{4}r^2\sigma
 \, (\rho^2\partial_\mu\xi-\epsilon_{\mu\nu}\partial^\nu\psi )^2}
 {(\rho^2 + \frac{1}{4}r^2\sigma^2)^2} + \frac{1}{2}( \rho^2+1)\sigma
 +2 \lambda \, r^2 \left(\sigma^2-1\right)\sigma
 - \rho \sigma\cos \xi = 0
\label{giSigma}
\end{equation}
\begin{equation}
 \partial^\mu\left\{ \frac{\partial_\mu \psi -  \rho^2
 \epsilon_{\mu\nu}\partial^\nu\xi}{\rho^2 +
 \frac{1}{4}r^2\sigma^2}\right\} + \frac{2 }{r^2} \,\psi = 0
\label{giPsi}
\end{equation}
\begin{equation}
 \partial^\mu\left\{ \frac{\rho^2(\frac{1}{4} r^2\sigma^2
 \partial_\mu\xi +\epsilon_{\mu\nu}\partial^\nu\psi)}{\rho^2
 + \frac{1}{4}r^2\sigma^2}\right\}+ \frac{1}{2}\rho\sigma^2
 \sin \xi= 0 \ ,
\label{giPsiXiOne}
\end{equation}
\label{giEqs}
\end{mathletters} \hskip -5pt
}
where the indices run over $0$ and $1$ and are raised and
lowered with the metric\footnote{The sign convention of the
metric in this paper is opposite to that of Ref.~\cite{gi}.}
\hbox{$\eta_{\mu\nu}= {\rm diag}(1,-1)$}, so that $\partial_\mu
\partial^\mu=\partial_t^2-\partial_r^2$.
The energy takes the form
 \begin{eqnarray}
 \label{giEnergy}
 \epsilon = 4\pi && \int_0^\infty dr \bigg\{ ~
 (\partial_t\rho)^2 + (\partial_r\rho)^2 +
 r^2 (\partial_t\sigma)^2 + r^2 (\partial_r\sigma)^2 +
 \nonumber \\  &&
 \frac{\frac{1}{4}r^2\sigma^2\rho^2}
 {\rho^2 + \frac{1}{4}r^2\sigma^2 }\left[ (\partial_t\xi)^2 +
 (\partial_r\xi)^2\right] + \frac{1} {\rho^2 +\frac{1}{4}r^2\sigma^2 }
 \left[ (\partial_t\psi)^2 + (\partial_r\psi)^2\right]+
 \nonumber \\ &&
 \frac{2\psi^2}{r^2} + \frac{1}{2 r^2}(\rho^2-1)^2 +
 \frac{1}{2}(\rho^2+1)\sigma^2 - \rho\sigma^2\cos \xi +
 \lambda\, r^2 \left(\sigma^2- 1\right)^2 ~ \bigg\} \ ,
 \end{eqnarray}
and we see that the vacuum is given by $\rho_{\rm vac}=1$,
$\sigma_{\rm vac}=1$, $\psi_{\rm vac}=0$ and $\xi_{\rm vac}=0$.

We wish to consider small fluctuations about the vacuum.
It is convenient to define shifted fields $y$ and $h$ by
\begin{mathletters}
\label{varShift}
\begin{eqnarray}
 \rho(r,t) &=& 1 + y(r,t) \\
 \label{rhoShift}
 \sigma(r,t) &=& 1 + \frac{h(r,t)}{r} \ .
\label{sigmaShift}
\end{eqnarray}
 \end{mathletters} \hskip -5pt
Then to linear order in $h$, $y$, $\psi$ and $\xi$,
(\ref{gieq}) becomes

\vbox{
\begin{mathletters}
\label{lineqns}
\begin{eqnarray}
 \Biggl(\partial_\mu\partial^\mu + 4\lambda \Biggr)\,h  &=&0
\label{giSigmaLin} \\
 \Biggl(\partial_\mu\partial^\mu + \frac{1}{2}  +
 \frac{2}{r^2}\Biggr)\,y &=& 0
\label{giRhoLin} \\
 \partial^\mu\left\{ \frac{\partial_\mu \psi -  \epsilon_{\mu\nu}
  \partial^\nu \xi}{1 + \frac{1}{4}r^2}\right\} + \frac{2}{r^2} \,
 \psi &=& 0
\label{giPsiLin} \\
 \partial^\mu\left\{ \frac{ \frac{1}{4} r^2 \partial_\mu \xi +
 \epsilon_{\mu\nu}\partial^\nu \psi }{1 + \frac{1}{4}r^2 }
 \right\} + \frac{1}{2} \xi &=& 0 \ .
\label{giXiLin}
\end{eqnarray}
\end{mathletters} \hskip -5pt
}
\noindent
Equation (\ref{giSigmaLin}) corresponds to a pure Higgs field
excitation characterized by mass \hbox{$m_H= 2 \sqrt{\lambda}$},
while (\ref{giRhoLin})-(\ref{giXiLin}) are the three gauge
modes of mass $m_W=1/\sqrt{2}$.\footnote{Upon restoring the
factors of $g$ and the Higgs field vacuum expectation value
$v$, these masses take the standard form
$m_H=\sqrt{2\lambda} \, v$ and $m_W=\frac{1}{2}g \, v$.}
To implement the boundary conditions (\ref{bczeror}), we take
the gauge invariant fields $h$, $y$, $\psi$ and $\xi$
to vanish at $r=0$. At $r=L$ we take $h$, $y$ and
$\xi$ to vanish (consistent with $\chi$ and $\phi$ taking
their vacuum values there). The $r=L$ boundary condition
on $\psi$ is that $\partial_r\psi$ is zero ($\psi$ can not
vanish at large $r$ since it is proportional to the time
derivative of the gauge field). We wish to solve (\ref{lineqns})
subject to these boundary conditions, and then
extract the corresponding amplitudes.

Let us examine the four types of modes in turn. They can all
be expressed in terms of the spherical Bessel functions
(\ref{sphbess}). Equation (\ref{giSigmaLin}) produces an
eigenmode whose non-vanishing components are of the
form $h_n(r,t)=h_n(r)\cos \omega_{1 n} t$, with
\begin{equation}
h_n(r)=\lambda_{1 n} r \, j_0(\lambda_{1 n} r) \, N_{1 n} \ ,
\label{heq}
\end{equation}
where $\omega_{1 n} =(4\lambda+\lambda_{1 n}^2)^{1/2}$ and
$\lambda_{1 n}=n\pi/L$ for $n=1,2,\cdots$. The parameters
$\lambda_{1 n}$ have been chosen so that $h_n(L,t)=0$, and
the normalization constants $N_{1 n}$ are taken to be
\begin{equation}
N_{1 n} = \left[ \, \frac{2}{L} \, \right ]^{1/2}
\label{Nonen}
\end{equation}
so that the $h_n(r)$ are orthonormal over the interval
$[0,L]$. To extract these modes from a given solution
we expand the Higgs excitation as
\begin{eqnarray}
h(r,t) &=& \sum_n A_n \, h_n(r) \cos \omega_{1 n} t
\end{eqnarray}
with
\begin{eqnarray}
A_n = \left\{
\left[ \int_0^L dr \, h(r,t) h_n(r)\right]^2 +\frac{1}{\omega_{1 n} ^2}
\left[ \int_0^L dr \, \dot h(r,t) h_n(r)\right]^2
\right\}^{1/2} \ ,
\label{Bone}
\end{eqnarray}
where the dot denotes the time derivative. To find the
associated amplitudes, we consider the energy of a pure
$h$-excitation. Using (\ref{giEnergy}), and the boundary
conditions on $h$, the quadratic energy is
\begin{eqnarray}
H_h = \int_0^L dr \, \left\{ (\partial_t h)^2
+(\partial_r h)^2 + 4\lambda h^2 \right\} \ .
\end{eqnarray}
Integrating the second term by parts and using the
equation of motion (\ref{giSigmaLin}) we find
\begin{mathletters}
\begin{eqnarray}
H_h &=& 4\pi \int_0^L dr \, \left\{ (\partial_t h)^2 -
h \,\partial_t^2 h
\right\}
\label{Hhlin} \\
&=& 4\pi \sum_n A_n^2 \, \omega_{1 n} ^2  \ .
\end{eqnarray}
\end{mathletters} \hskip -5pt
Hence, the modulus squared of the amplitudes
for this first mode are
\begin{eqnarray}
|a_{1 n}|^2 = 4\pi \,A_n^2 \, \omega_{1 n}  \ ,
\end{eqnarray}
where $\omega_{1 n} = [4\lambda + (x_{1 n} /L)^2]^{1/2}$,
$x_{1 n}=n \pi$ with $n=1,2,\cdots$
and the $A_n $ are given by (\ref{Bone}).

Equation (\ref{giRhoLin}) produces an eigenmode whose
non-vanishing components are of the form $y_n(r,t)=y_n(r)
\cos \omega_{2 n} t$, with
\begin{equation}
y_n(r)=\lambda_{2 n} r \, j_1(\lambda_{2 n} r) \, N_{2 n} \ ,
\label{yeq}
\end{equation}
where $\omega_{2 n} =(1/2+\lambda_{2 n}^2)^{1/2}$ and
$\lambda_{2 n} \equiv x_{2 n}/L$, with $x_{2 n}$ being
the positive solutions to $\tan x_{2 n} = x_{2 n}$
(with this set of modes and those that follow, we will label the
normal modes starting from $n=1)$. The
parameters $\lambda_{2 n}$ have been chosen so that
$y_n(L,t)=0$, and the normalization constants $N_{2 n}$
are taken to be
\begin{equation}
N_{2 n} = \left[\, \frac{2}{L \sin^2 x_{2 n}}\,\right]^{1/2}
\label{Ntwon}
\end{equation}
so that the $y_{n}(r)$ are orthonormal over $[0,L]$.
To extract the amplitudes from a given
solution we first expand the $y$-excitation as
\begin{eqnarray}
y(r,t) &=& \sum_n B_n \, y_n(r) \cos \omega_{2 n} t
\end{eqnarray}
with
\begin{eqnarray}
B_n = \left\{
\left[ \int_0^L dr \, y(r,t) y_n(r)\right]^2 +\frac{1}{\omega_{2 n} ^2}
\left[ \int_0^L dr \, \dot y(r,t) y_n(r)\right]^2
\right\}^{1/2} \ .
\label{Btwo}
\end{eqnarray}
Using (\ref{giEnergy}), the quadratic energy of a pure
$y$-excitation is
\begin{eqnarray}
H_y = \int_0^L dr \, \left\{ (\partial_t y)^2
+(\partial_r y)^2 + \frac{y^2}{2} +
\frac{2 y^2}{r^2} \right\} \ .
\end{eqnarray}
Integrating the second term by parts and using the
equation of motion (\ref{giRhoLin}) we find
\begin{mathletters}
\begin{eqnarray}
H_y &=& 4\pi \int_0^L dr \, \left\{ (\partial_t y)^2 -
y \,\partial_t^2 y
\right\}
\label{Hylin} \\
&=& 4\pi \sum_n B_n^2 \, \omega_{2 n} ^2 
\end{eqnarray}
\end{mathletters} \hskip -5pt
Hence, the modulus squared of the amplitudes
for the second mode are
\begin{eqnarray}
|a_{2 n}|^2 = 4\pi \,B_n^2 \, \omega_{2 n}  \,
\end{eqnarray}
where $\omega_{2 n} = [1/2 + (x_{2 n}/L)^2]^{1/2}$, with
$x_{2 n}$ being the positive solutions of $\tan x_{2 n}=
x_{2 n}$, and where the $B_n $ are given by (\ref{Btwo}).

The remaining two modes are more involved since (\ref{giPsiLin})
and (\ref{giXiLin}) are two coupled equations for $\psi$ and $\xi$.
To disentangle these modes, we first rewrite these equations as
\begin{mathletters}
\label{xipsiab}
\begin{eqnarray}
\partial_t^2\psi -\partial_r^2\psi + \frac{\psi}{2} + \frac{2\psi}{r^2} +
\frac{2 r}{4+r^2}\left[ \partial_r\psi + \partial_t \xi \right] &=& 0
\label{xipsia} \\
\partial_t^2\xi-\partial_r^2\xi + \frac{\xi}{2} + \frac{2\xi}{r^2} -
\frac{8}{r(4+r^2)}\left[ \partial_r\xi + \partial_t \psi \right] &=& 0 \ .
\label{xipsib}
\end{eqnarray}
\end{mathletters} \hskip -5pt
We now define $\zeta=r (\partial_r\psi+\partial_t\xi)/(4+r^2)$, so that
(\ref{xipsiab}) may be rewritten as
\begin{mathletters}
\begin{eqnarray}
\partial_t^2\psi -\partial_r^2\psi + \frac{\psi}{2} + \frac{2\psi}{r^2} +
2 \zeta &=& 0
\label{psiya} \\
\partial_t^2\zeta -\partial_r^2\zeta + \frac{\zeta}{2} +
\frac{2 \zeta}{r^2} &=& 0 \ .
\label{psiyb}
\end{eqnarray}
\end{mathletters} \hskip -5pt
Equation (\ref{psiya}) follows directly from (\ref{xipsia}) and
the definition of $\zeta$, while (\ref{psiyb}) is derived as
follows. First, take a time derivative of (\ref{xipsib}). This
gives a $\partial_t^2\psi$ term in the square brackets, which
may be eliminated using (\ref{psiya}) to give
\begin{eqnarray}
 \partial_t^2\dot\xi - \partial_r^2 \dot\xi + \frac{\dot\xi}{2} +
 \frac{2 \dot\xi}{r^2} - \frac{8}{r(4+r^2)}\left[\partial_r\dot\xi
+ \partial_r\psi' -2\zeta \right] + \frac{4\psi}{r^3} = 0 \ ,
\label{dotxi}
\end{eqnarray}
where the dot and prime denote time and space derivatives
respectively. We have written $\partial_t^2\dot\xi$ rather than
$\partial_t^3\xi$, $\partial_r\psi'$ rather than $\partial_r^2\psi$,
etc. for future convenience. Taking a spatial derivative
of (\ref{psiya}) gives
\begin{eqnarray}
\partial_t^2\psi' -\partial_r^2\psi' + \frac{\psi'}{2} +
 \frac{2\psi'}{r^2}
- \frac{4\psi}{r^3} + 2 \zeta' &=& 0 \ .
\label{primepsi}
\end{eqnarray}
Adding (\ref{dotxi}) and (\ref{primepsi}), and using
$\dot\xi+\psi'=(4+r^2)\zeta/r$ gives (\ref{psiyb}).

These normal modes fall into two classes, one in which
$\zeta=0$ and another in which $\zeta$ is non-vanishing.
In the former case, (\ref{psiya}) may be solved for $\psi$.
We may then use \hbox{$\partial_t\xi + \partial_r\psi=0$}
to solve for $\xi$. Thus, mode three takes the form
$\psi_{3 n}(r,t)=\psi_{3 n}(r)\sin \omega_{3 n} t$ and
$\xi_{3 n}(r,t)=\xi_{3 n}(r)\cos \omega_{3 n} t$,
and after some algebra we find
\begin{mathletters}
\begin{eqnarray}
\psi_{3 n}(r) &=& \lambda_{3 n} r \, j_1(\lambda_{3 n} r)
\, N_{3 n}  \\
\xi_{3 n}(r) &=& \frac{\lambda_{3 n}}{\omega_{3 n}}
\left[ 2 j_1(\lambda_{3 n} r) -
\lambda_{3 n} r \, j_2(\lambda_{3 n} r)\right]\, N_{3 n}  \ ,
\end{eqnarray}
\end{mathletters} \hskip -5pt
where $\omega_{3 n}=(1/2+\lambda_{3 n}^2)^{1/2}$
and $\lambda_{3 n}\equiv x_{3 n}/L$, with $x_{3 n}$
being the positive solutions to \hbox{$\tan x_{3 n} =
x_{3 n}/(1-x_{3 n}^2)$}. The parameters $\lambda_{3 n}$
have been chosen so that $\xi_{3 n}(L,t)=0$ (since
$\zeta$ vanishes, this automatically ensures that
$\partial_r \psi_{3n}(L,t)=0$). The normalization
constants $N_{3 n}$ will be chosen below to ensure
a convenient orthonormality relation for the $\psi_{3 n}(r)$
and $\xi_{3 n}(r)$.

We turn now to the other class of modes in which $\zeta$ is
non-vanishing. We can first solve (\ref{psiyb}) for $\zeta$,
and then solve (\ref{psiya}) for $\psi$ treating $\zeta$ as
a source. Then, using the definition of $\zeta$, we can
solve for $\xi$. Again, writing $\psi_{4 n}(r,t)= \psi_{4 n}(r)
\sin \omega_{4 n} t$ and $\xi_{4 n}(r,t)=\xi_{4 n}(r)
\cos \omega_{4 n} t$, we find
\begin{mathletters}
\begin{eqnarray}
\psi_{4 n}(r) &=& \frac{r}{\lambda_{4 n}^2}
\left[ 2 j_1( \lambda_{4 n} r) -
\lambda_{4 n} r \, j_0(\lambda_{4 n} r)\right]
\, N_{4 n} \\
\xi_{4 n}(r) &=& \frac{1}{\lambda_{4 n}^2 \omega_{4 n}} \left[ -2
\lambda_{4 n} r \,j_0(\lambda_{4 n} r) + 4(1-\lambda_{4 n}^2)
j_1(\lambda_{4 n} r) - 2 \lambda_{4 n} r \, j_2(\lambda_{4 n} r)
\right] \, N_{4 n} \,
\end{eqnarray}
\end{mathletters} \hskip -5pt
where $\omega_{4 n}=(1/2+\lambda_{4 n}^2)^{1/2}$ and
$\lambda_{4 n}\equiv x_{4 n}/L$, with $x_{4 n}$ being the
positive solutions to \hbox{$\tan x_{4 n} = x_{4 n}$}. The
parameters $\lambda_{4 n}$ have been chosen so that
$\xi_{4 n}(L,t)=0$ and \hbox{$\partial_r \psi_{4n}(L,t)=0$},
and the normalization constants $N_{4 n}$ will be chosen below.

We expand a $\psi$-$\xi$ excitation as
\begin{mathletters}
\label{psixiex}
\begin{eqnarray}
\psi(r,t) &=& \sum_{j=3,4}
\sum_n C_{j n} \, \psi_{j n}(r) \sin \omega_{j n} t \\
\xi(r,t) &=& \sum_{j=3,4}
\sum_n C_{j n} \, \xi_{j n}(r) \cos \omega_{j n} t \ .
\end{eqnarray}
\end{mathletters} \hskip -5pt
Choosing the normalization constants
\begin{mathletters}
\begin{eqnarray}
 N_{3 n} &=& \left[ \,\frac{(2-x_{3 n}^2)\sin^2x_{3 n}}{2 L x^4_{3 n}}
 + \frac{x_{3 n}^4-x_{3 n}^2-2}{2 L x_{3 n}^2} \, \right]^{-1/2}
\\
 N_{4 n} &=& \left[ \,\frac{L^3(2 x_{4n}^6+L^2)\sin^2 x_{4 n}}
 {x_{4 n}^8} + \frac{L^5 (x_{4 n}^2-1)}{x_{4n}^6} \, \right]^{-1/2} \ ,
\end{eqnarray}
\end{mathletters} \hskip -5pt
the modes satisfy the orthonormality relations
\begin{mathletters}
\label{orthorel}
\begin{eqnarray}
 \int_0^L dr \, \Bigg\{ \,
 \frac{r^2}{4+r^2} \, \omega_{j n}\xi_{j n}(r) \,
 \omega_{k m}\xi_{k m}(r)
 + \frac{4}{4+r^2} &&
 \, \partial_r\psi_{j n}(r)\,\partial_r\psi_{k m}(r)
\nonumber \\ &&
 +  \frac{2}{r^2} \, \psi_{j n}(r)\psi_{k m}(r) \Bigg\} =
  \delta_{n m} \delta_{j k}
\\
 \int_0^L dr \, \Bigg\{ \, \frac{4}{4+r^2} \,
 \omega_{j n}\psi_{j n}(r) \, \omega_{k m}\psi_{k m}(r)
 + \frac{r^2}{4+r^2} && \, \partial_r\xi_{j n}(r)\,
 \partial_r\xi_{k m}(r)
\nonumber \\ &&
 + \frac{1}{2} \, \xi_{j n}(r)\xi_{k m}(r)
 \Bigg\} \hskip0.1in =  \delta_{n m} \delta_{j k} \ .
\end{eqnarray}
\end{mathletters} \hskip -5pt
Using (\ref{orthorel}) in (\ref{psixiex}), the overlap
coefficients $C_{j n}$ become
\begin{eqnarray}
 C_{j n}  = && \Bigg\{ \Bigg[
 \int_0^L dr \Bigg( - \omega_{j n} \frac{r^2}{4+r^2}\,
 \partial_t \xi(r,t) \,\xi_{j n}(r)
\nonumber \\ &&
 + \frac{4}{4+r^2} \partial_r\psi(r,t) \, \partial_r\psi_{j n}(r) +
 \frac{2}{r^2}\, \psi(r,t) \psi_{j n}(r) \Bigg) \Bigg]^2
\nonumber \\ &&
 + \Bigg[ \int_0^L dr \Bigg(\omega_{j n} \frac{4}{4+r^2}\,
 \partial_t \psi(r,t) \, \psi_{j n}(r)
\nonumber \\ &&
 + \frac{r^2}{4+r^2}\,
 \partial_r\xi(r,t) \, \partial_r\xi_{j n}(r) +
 \frac{1}{2}\, \xi(r,t) \xi_{j n}(r) \Bigg) \Bigg]^2
 \Bigg\}^{1/2} \ .
\end{eqnarray}
To extract the amplitudes, consider a pure $\psi$-$\xi$
excitation. Using (\ref{giEnergy}), the quadratic energy
is given by
\begin{mathletters}
\begin{eqnarray}
H_{\xi\psi} &=& 4\pi \int_0^L dr \, \left\{ \frac{r^2}{4+r^2}
\left[ \dot\xi^2 +\xi'^2\right] + \frac{4}{4+r^2}
\left[ \dot\psi^2 +\psi'^2\right] + \frac{2\psi^2}{r^2} +
\frac{\xi^2}{2} \right\} \\
&=& 4\pi \sum_{j=3,4}\sum_n C_{j n}^2  \ ,
\end{eqnarray}
\end{mathletters} \hskip -5pt
hence
\begin{eqnarray}
|a_{j n}|^2 = 4\pi \, \frac{C_{j n}^2 }{\omega_{j n}} \qquad j=3,4 \ .
\end{eqnarray}

Even though we have solved the normal modes problem
analytically in the continuum, the amplitudes $|a_{j n}|^2$
will be extracted using discrete numerical solutions.
This is justified by the large size of our lattice: $N=2239$,
$\Delta r = 0.04$ (with $\lambda=0.1$).

For computational purposes it is important to note that,
strictly speaking, completeness sums involve all normal
modes, but in a physically meaningful situation they will
be saturated well before the normal mode indices reach
the maximum value $N$. The highest normal modes
indeed correspond to artifacts of the discretization.
Thus, to avoid unnecessary computational burdens, we
will place a cutoff ${N_{\rm mode}}\sim N/5$ to $N/10$
on the number of normal modes and calculate the Higgs
and gauge boson particle numbers as
\begin{mathletters}
\label{nuhg}
\begin{eqnarray}
\nu_{\rm higgs} &=& \sum_{n=1}^{N_{\rm mode}}
|a_{1n}|^2  \\
\nu_{\rm gauge} &=& \sum_{n=1}^{N_{\rm mode}}
\left\{ |a_{2n}|^2 +
|a_{3n}|^2 +|a_{4n}|^2 \right\} \ .
\end{eqnarray}
\end{mathletters} \hskip -4pt
The total particle number is given by
\begin{eqnarray}
\nu = \nu_{\rm higgs} + \nu_{\rm gauge} \ .
\end{eqnarray}
We have verified that our results are insensitive to this cutoff,
which means that short wavelength modes comparable to the lattice
spacing are not excited in any appreciable manner. One should
also note that our procedure for calculating the particle number is
obviously gauge invariant (as it should be) since it makes use of an
expansion into normal modes of gauge invariant variables.

\vskip0.5in

\vbox{
\begin{figure}
\centerline{
\epsfxsize=100mm
\epsfbox{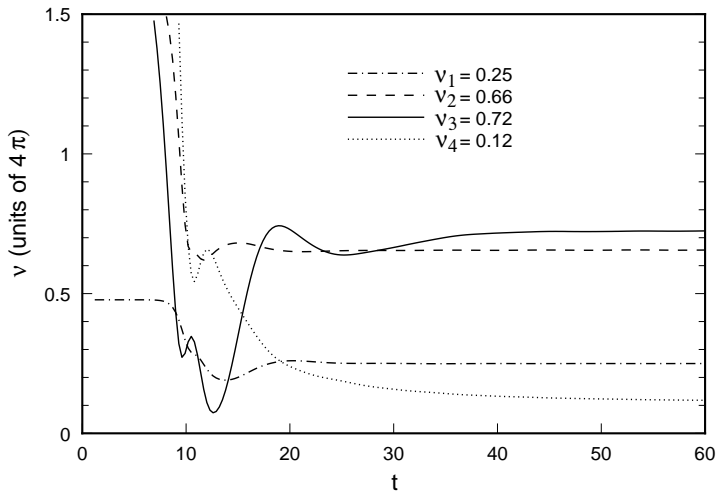}
}
\caption{\tenrm
Decay of a small perturbation about the sphaleron:
behavior of the particle number in the four normal
modes of oscillation of the linearized system as function
of time for lattice parameters $N=2239$, $\Delta r = 0.04$
and $N_{\rm mode}=200$ with $\lambda=0.1$. The physical
particle numbers are obtained by multiplying the asymptotic
values in the graph by $4\pi/g^2 \sim 30$, which gives
$N_{\rm higgs}\sim8$ and $N_{\rm gauge}\sim45$, for a total
physical particle number of $N_{\rm phys}\sim53$.
}
\end{figure}
}

In Fig.~7 we display the behavior of the particle number in
the four normal modes of oscillation as function of time. The
initial state is the small perturbation about the sphaleron in
Fig.~5, which gives rise to outgoing spherical waves as the
configuration decays. This is the state from which we start the
stochastic sampling procedure described in the next section.
Since the energy density is distributed over an expanding
shell, the system quickly approaches the linear regime.
This is apparent from Fig.~7 where, after an initial transition
period in which the particle numbers of the four modes are not
constant, they settle to values which are reasonably
constant in time. We take this as evidence that the system
has indeed reached an asymptotic linear regime where
one can define a conserved particle number.

There are two additional quantities that are useful in
measuring the extent of linearity, namely the spectral
energy $\epsilon_{\rm spec}$ and the linearized energy
$\epsilon_{\rm lin}$. The spectral energy is defined as the
sum over normal mode energies,
\begin{eqnarray}
\epsilon_{\rm spec} = \sum_{n=1}^{N_{\rm mode}}
\left\{ \omega_{1 n}|a_{1n}|^2 + \omega_{2 n}|a_{2n}|^2 +
\omega_{3 n} |a_{3n}|^2 + \omega_{4 n}|a_{4n}|^2 \right\} \ ,
\end{eqnarray}
while the linearized energy is defined by integrating the energy
density in (\ref{giEnergy}) expanded to second order in a
perturbation about the vacuum,
 \begin{eqnarray}
 \label{epslindef}
 \epsilon_{\rm lin} &=& 4\pi \int_0^\infty dr \bigg\{ ~
 (\partial_t y)^2 + (\partial_r y)^2 + \frac{2 y^2}{r^2} +
 \frac{y^2}{2} +
 (\partial_t h)^2 + (\partial_r h)^2 + 4\lambda h^2 +
\nonumber \\  &&
 \frac{r^2} {4+r^2 }\left[ (\partial_t\xi)^2 +
 (\partial_r\xi)^2\right] + \frac{\xi^2}{2} +
 \frac{4} {4+r^2 }
 \left[ (\partial_t\psi)^2 + (\partial_r\psi)^2\right]+
 \frac{2\psi^2}{r^2} \bigg\} \ .
 \nonumber
\end{eqnarray}
%
Both the spectral and linear energy are gauge invariant since
they have been defined using gauge invariant quantities. If
the system linearizes, then both $\epsilon_{\rm spec}$
and $\epsilon_{\rm lin}$ should be close to the conserved
total energy $\epsilon$, which is given by the integral
(\ref{giEnergy}) (or in terms of gauge-variant variables by
(\ref{HpHC})). The total energy of the configuration in Fig.~7
is given by \hbox{$\epsilon / 4 \pi = 2.5447$}, while the
asymptotic spectral and linear energies are given by
\hbox{$\epsilon_{\rm spec} / 4 \pi = 2.5679$} and
\hbox{$\epsilon_{\rm lin} / 4 \pi = 2.5685$},
and we see that the system has linearized to within one
percent. (We also see that the sum over the energies of
individual modes, although cut off at $N_{\rm mode}$,
essentially accounts for all the linearized energy.)

\vbox{
\begin{figure}[t]
\centerline{
\epsfxsize=100mm
\epsfbox{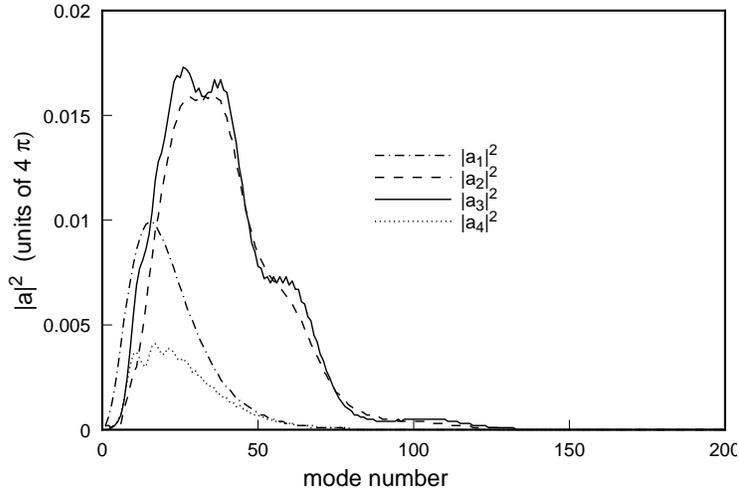}
}
\caption{\tenrm
Mode distribution for the asymptotic state of Figs.~5
and~7. This distribution is gauge invariant, and shows
that all the particle are rather soft and comparable in
energy.
}
\end{figure}
}

We can also investigate the mode distribution by examining
the amplitudes $|a_{j n}|^2$ as a function of mode number $n$.
As the system linearizes and the particle number becomes well
defined, the mode distribution also becomes constant in time.
Figure~8 illustrates the distribution of the asymptotic linear
state of Figs.~5~and~7. Note that the population of the system
is heavily weighted towards low lying modes. The mode
cutoff used in calculating
the particle number was $N_{\rm mode}=200$, and we see
that modes greater than about $n=150$ are not populated to
any appreciable extent. The mode distributions are heavily
peaked near \hbox{$n_{\rm pk} \sim 50$}, which corresponds
to a frequency of \hbox{$\omega_{\rm pk} \sim n_{\rm pk}\pi/
L \sim 0.1$}. The perturbation about the sphaleron of
Figs.~5~and~7 decays into about $50$ rather soft particles
(in physical units), each one of comparable energy. Finally,
we point out that the mode distribution is gauge invariant as well.

\section{Stochastic Sampling of Initial Configurations}
\hspace{1cm}
As we have discussed, our goal is to find the region in the
\hbox{$\epsilon$-$\nu$} plane spanned by all topology
changing classical solutions. More specifically, we would
like to find the lower boundary of this region. The tools
we have at our disposal allow us to vary the coefficients
$c_n$ of (\ref{fieldparam}),
which defines the system as it passes over the sphaleron barrier,
and to calculate the corresponding energy $\epsilon$ and
incoming particle number $\nu$. From the computational point
of view, $\epsilon$ and $\nu$ can be considered as known
functions (albeit laboriously obtained) of the variables $c_n$.
We would then like to find
\begin{equation}
\nu_{\rm lower}(\epsilon) = {\rm Min}_{\{c {\scriptscriptstyle n}, \,
{\rm fixed} \; \epsilon \}}\; \nu \ .
\label{condmin}
\end{equation}

The particle number $\nu$ may have several local minima since it
is a highly non-linear function of the variables $c_n$, and
a straightforward constrained minimization procedure, such
as a conjugate gradient technique, could fail to reveal the
absolute minimum of $\nu$ at a given $\epsilon$. We therefore
decided to solve the problem using stochastic sampling.
Stochastic sampling methods, driven by suitable weight
functions and in combination with annealing techniques, have
indeed proven very effective in exploring the overall structure of
complicated surfaces and in approximating their global
minima.

Our procedure consists in generating ``configurations'' of the
system weighted by a function
\begin{equation}
W = \exp(- F) \ ,
\label{weighta}
\end{equation}
with
\begin{equation}
F = \beta \, \epsilon \, - \mu \, \nu \ .
\label{weightb}
\end{equation}
By ``configuration'' simply we mean the collection of
variables $c_n$, which determine the whole evolution
of the system. Since $\epsilon$ and $\nu$ are
functions of $c_n$, the weight given by (\ref{weighta})
and (\ref{weightb}) is also a function of $c_n$ and
defines a probability distribution
\begin{equation}
dP = Z^{-1} \, \prod_n d c_n \, W(c_n) \ .
\label{measure}
\end{equation}
We will generate topology changing configurations distributed
according to (\ref{measure}). Clearly, by taking large values
for the parameters $\beta$ and $\mu$ we will drive the distribution
strongly towards the lower boundary in the space of all topology
changing solutions. By using different ratios $\mu / \beta$ we
will be able to drive the distribution in different direction and
thus follow the lower envelope of the region, while temporarily
lowering the values of $\mu$ and/or $\beta$ will allow us to
anneal the distribution. We will typically take $\beta$ between
$50$ and $1,000$ while $\mu$ will range between $1,000$ and
$20,000$.

To generate the desired distribution we have used a
Metropolis Monte Carlo algorithm. Starting from a definite
configuration $c_n$, we randomly select one of the variables
$c_i$ and perform a variation $c_i \to c_i'=c_i + \Delta c_i$
(in our computation, the $ \Delta c_i$ are Gaussian distributed
with a mean of 0.0008.)
The system is evolved backward and forward in time and we
calculate the energy, in-state particle number and change
of winding number. If the winding number does not change,
we proceed to vary another of the variables $c_n$. If the
topology changes, we evaluate $\Delta F=\beta \Delta
\epsilon + \mu \Delta \nu$ and the new value $c_i'$ is
accepted with conditional probability $ p = {\rm Min}
[1, \exp (- \Delta F)]$. Specifically, we generate a pseudorandom
number $r$ uniformly distributed between $0$ and $1$, and if $r
\le \exp (- \Delta F)$ the change is accepted and the new value
$c_i'$ replaces the old one. Otherwise, if $r >
\exp (- \Delta F)$ the old value is kept and we select another
of the variables $c_n$ for a possible upgrade. (We should note
here that when the winding number changes, even if the trial
value $c_i'$ is rejected, we still record its value and the
corresponding values of $\epsilon$ and $\nu$, since they do
correspond to a possible topology changing evolution.)

It must be emphasized that although our algorithm generates
a distribution of topology changing solutions of the equations
of motion, this distribution represents only a computational
device and carries no special physical significance. Indeed,
the probability measure (\ref{measure}) is based on the arbitrary
choice of variables $c_n$ and no Jacobian factor of any kind
has been introduced. It would be possible to define a measure
which represents a physically meaningful distribution, and our
notation $\beta$ and $\mu$ for the weights of $\epsilon$ and
$\nu$ has been inspired by the analogy with a grand canonical
ensemble. But still, in the present context, there is no
reason for defining any particular physically meaningful measure
and no justification for the attached computational costs.

\vbox{
\begin{figure}
\centerline{
\epsfxsize=100mm
\epsfbox{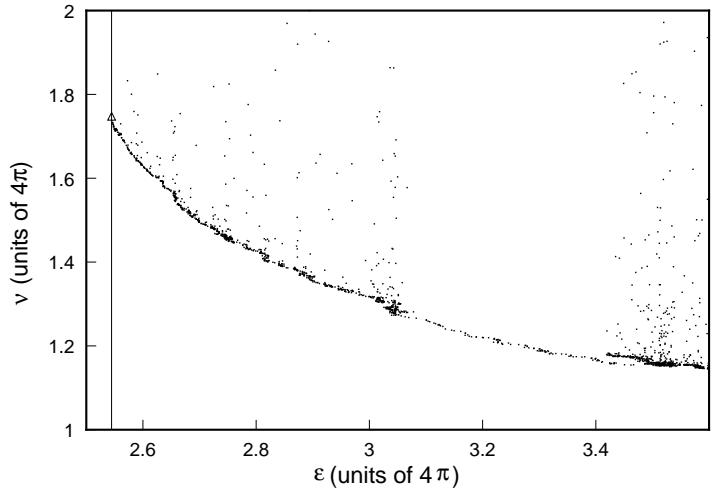}
}
\caption{\tenrm
Monte Carlo results with lattice parameters of $N=2239$,
$\Delta r = 0.04$ (giving L=89.56), $N_{\rm mode}=200$
and $N_{\rm sph}=50$, and with a Higgs self-coupling of
$\lambda=0.1$. The solid line marks the sphaleron
energy $\epsilon_{\rm sph}= 4\pi (2.5426)$, below which
no topology changing process can lie. The diamond represents
the configuration from which we seeded our Monte Carlo
search. To obtain quantities in physical units, multiply the
numbers along the axes by $4\pi/g^2 \sim 30$. The energy axis
extends from about $10 {~\rm TeV}$ to $15 {~\rm TeV}$, while the
particle number axis ranges from about $30$ particles to $60$.
}
\end{figure}
}

Figure~9 illustrates the results of our Monte Carlo
investigation. It represents about 300 hours of CPU time on
a 16 node partition of a CM-5. We generated approximately 30,000
configurations of which approximately 3,000 representatives
are plotted in the figure. We have chosen lattice parameters
$N=2239$ and $\Delta r = 0.04$, for a lattice of size $L=89.56$.
We have used a cutoff $N_{\rm mode}=200$ on the sums over
the modes, and the dimension of the initial configuration space
over which we have sampled is determined by $N_{\rm sph}=50$.
We have taken the Higgs self-coupling to be $\lambda=0.1$,
which in lattice units corresponds to a mass of about
$m_H=(40\Delta r)^{-1}$, or a physical mass of $m_H =110
{~\rm GeV}$. As one can see, our lattice is sufficiently dense that
there are many lattice sites within a single Higgs Compton
wavelength.

It is apparent from Fig.~9 that our search procedure is effective
in reducing the particle number and in exploring the lower
boundary of the space of topology changing classical solutions.
The complex nature of this space (or at least of our search
procedure) is also apparent from the figure, in that one can
clearly observe two breaks in the outline of the lower boundary
at \hbox{$\epsilon/4\pi \sim 3$} and \hbox{$\epsilon/4\pi \sim
3.4$}. The reason for the discontinuity is that in a first
extended search we did not verify that every individual solution
changed topology (performing this check is costly in computer time),
trusting that topology change would be the typical outcome of an
evolution which passes over the sphaleron barrier. A subsequent
analysis revealed however that for a whole subset of our
configurations, comprised between \hbox{$\epsilon/4\pi \sim 3$}
and \hbox{$\epsilon/4\pi \sim 3.4$}, the topology did not change:
the system went over the sphaleron barrier a second time in the
reversed direction and returned to the original topological
sector. We discarded all these configurations and verified that
the topology changed in all the remaining ones. We then
implemented the check for topology change at every Monte Carlo
step and restarted our sampling procedure by annealing a topology
changing configuration obtained for \hbox{$\epsilon/4\pi \sim 3$}.
This second search produced the set of configurations which stand
out at slightly lower $\nu$ between \hbox{$\epsilon/4\pi \sim 3$}
and \hbox{$\epsilon/4\pi \sim 3.4$}.

\vbox{
\begin{figure}
\centerline{
\epsfxsize=80mm
\epsfbox{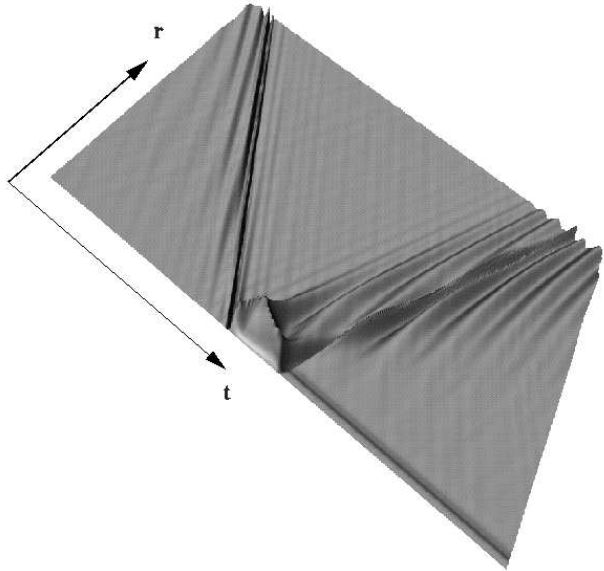}
}
\caption{\tenrm
Topology changing transition obtained after many Monte Carlo
iterations: behavior of the $\chi$ field. For a full color figure
see http://cthulu.bu.edu/$\scriptstyle\sim$bobs/bviolate.html.
}
\end{figure}
}

Our search procedure not only leads to classical solutions with
lower particle number, but is effective in selecting
configurations with special properties in the in-state (these two
features of course go hand in hand). In Fig.~10 we illustrate
the entire evolution for one of the topology changing processes
with low particle number, corresponding to one of the points at
the bottom-right corner of the plot in Fig.~9. Figure~10 should
be contrasted with Fig.~6 in which the evolution of our Monte
Carlo seed configuration is illustrated. The change is dramatic.
The in- and out-states in Fig.~6 look rather symmetric, up to the
topology change of the out-state concentrated about the origin.
The particle numbers of the in- and out-states are about the same
and of order $50$ in physical units ($\nu_i / 4\pi = 1.747$ and
$\nu_o / 4\pi = 1.750$, respectively). Figure~9 shows
that after many Monte Carlo iterations we have managed
to filter initial configurations $c_n$ so that the in-state
particle numbers are about $40\%$ lower
($\nu_i / 4\pi \sim 1.10$), and from Fig.~10 it is apparent
that the in-states are now much different from the out-states.
The former are narrow with the spectrum shifted towards
shorter wavelengths, while the outgoing states still display
the broad long-range waves seen at both ends of the
evolution in Fig.~6. Indeed, the particle number in the
out-state remains high.

\vbox{
\begin{figure}
\centerline{
\epsfxsize=100mm
\epsfbox{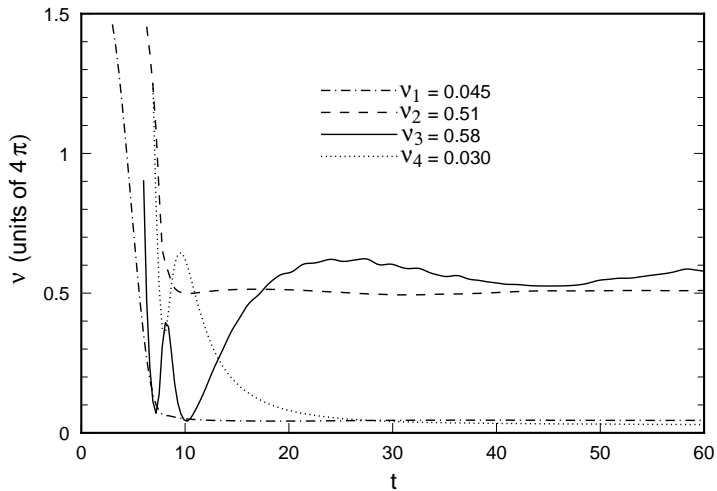}
}
\caption{\tenrm
Behavior of the in-state particle number in the four normal
modes. The initial state was obtained after many Monte Carlo
iterations, and soon linearizes. Note, however, that mode 3
remains about 10\% non-linear.
}
\end{figure}
}

More details of the configurations selected by our sampling
procedure are revealed by Figs.~11 and~12. Figure~11
illustrates the behavior of the particle numbers associated
with the four normal modes for the initial configuration $c_n$
used to generate Fig.~10. The asymptotic particle numbers
in Fig.~11 are associated with the in-state of the physically
relevant time reversed solution of Fig.~10. Figure~12
illustrates the mode distribution of this in-state. These
figures should be contrasted with Figs.~7~and~8 which
display the same quantities at the beginning of our search. The
change is again very impressive. In particular, it is clear that
the stochastic sampling procedure has selected classical
solutions where the mode distribution in the in-states is shifted
towards higher frequencies and shorter wavelengths. Of course,
this is necessary for a reduction of the ratio $\nu / \epsilon$.

\vbox{
\begin{figure}[t]
\centerline{
\epsfxsize=80mm
\epsfbox{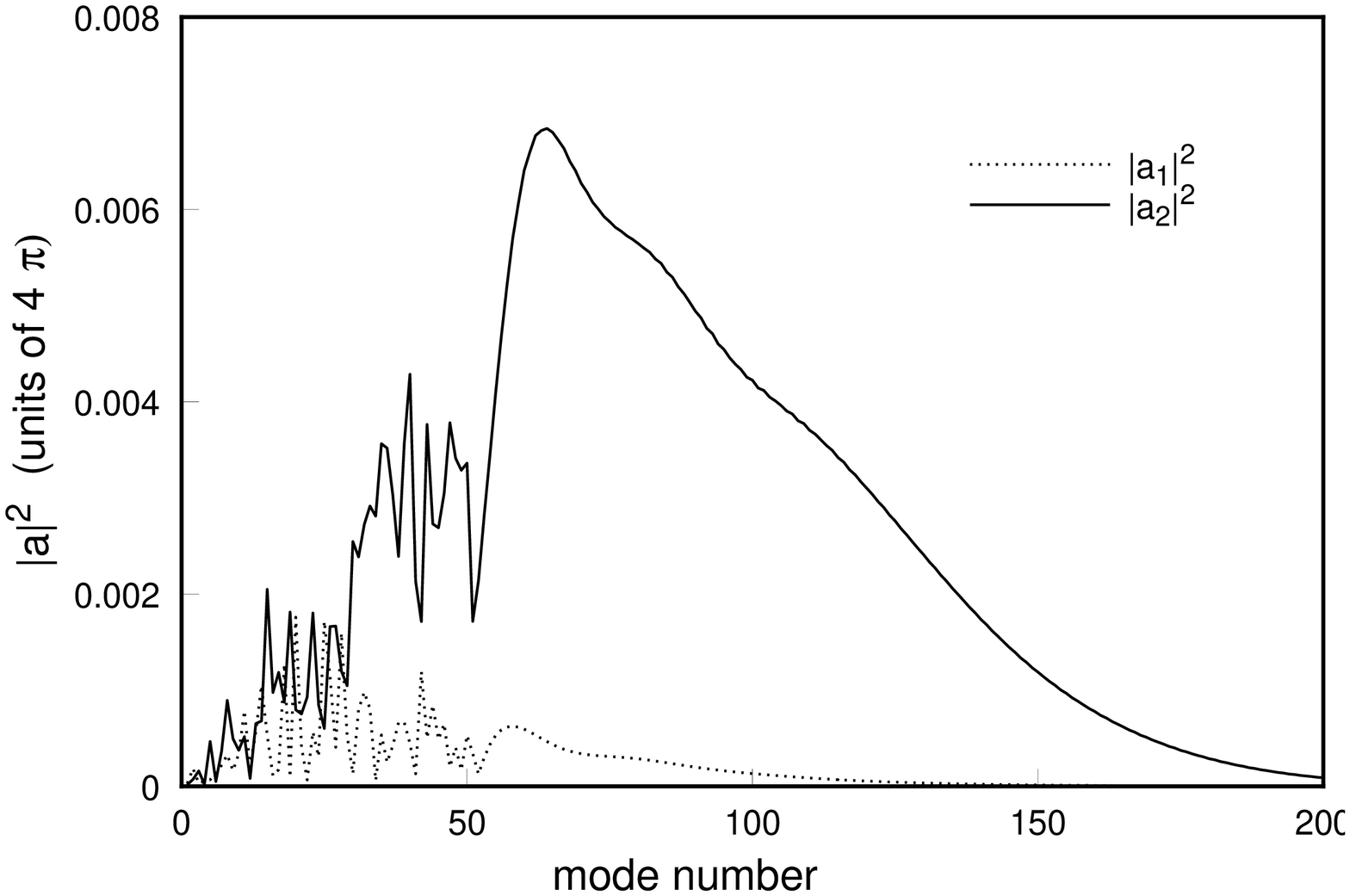}
}
\end{figure}

\begin{figure}
\centerline{
\epsfxsize=80mm
\epsfbox{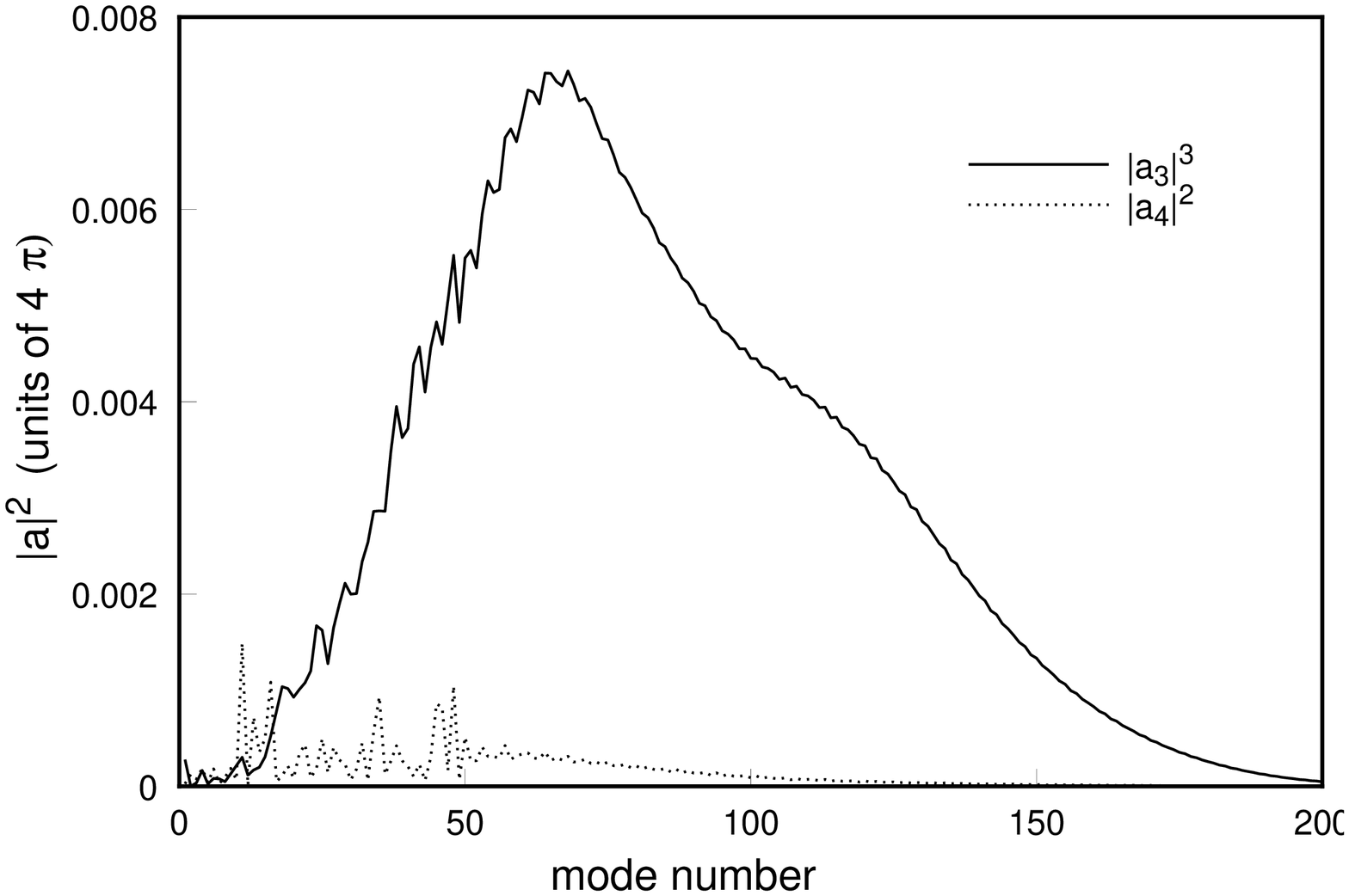}
}
\caption{\tenrm
Mode distribution of the asymptotic in-state of Fig.~11.
}
\end{figure}
}

Although our results show a marked decrease in the particle
number of the incoming state, nowhere in the energy range we have
explored does $\nu $ drop below $4\pi$, or in physical units
$N_{\rm phys}
{\ \lower-1.2pt\vbox{\hbox{\rlap{$>$}\lower5pt
\vbox{\hbox{$\sim$}}}}\ }30$ for
$E {\ \lower-1.2pt\vbox{\hbox{\rlap{$<$}\lower5pt
\vbox{\hbox{$\sim$}}}}\ } 15 {~\rm TeV}$.
This is a far cry from the value $N_{\rm phys} =2$
which would be needed to argue that baryon number
violation can occur in a high energy collisions. {}From
this point of view our present results are limited and
should be pushed to much higher values of $\epsilon$.
In the next section we will make some comments about
our future plans to explore higher energies and discuss
other investigations which can shed further light on
the properties of the system. As of now the computational
resources at our disposal, together with the rather
ambitious number of points we have used for our
numerical study, have not permitted us to go beyond
the energy range we have explored. We believe that
our results, as well as the formalism we have established,
are nevertheless interesting enough to warrant publication.
In some respect, the choice of a number of points as large
as our current $N = 2239$ has been an error of strategy.
In a preliminary investigation, described in \cite{rs}, we
had used $N=256$. The number of points in the lattice
determines of course the ultraviolet cutoff and this in turn
implies a minimum value for the ratio $\nu / \epsilon$.
This quantity is indeed minimized by placing all the
weight in the highest mode $N_{\rm mode}$, giving
$(\nu/\epsilon)_{\rm min} = 1/\omega_{\rm mode} \sim
L/N_{\rm mode}\pi$. With $N=256$ points we saw the onset of this
constraint, and we decided to chose a lattice size that would
push the lower limit on $\nu / \epsilon$ to a much smaller value
closer to the physically relevant domain. With the parameters of
our present calculation, the minimum would occur at $(\nu /
\epsilon)_{\rm min} \sim 0.15$. However, the increased
computational burden, together with the fact that the stochastic
sampling moved in the \hbox{$\epsilon$-$\nu$} plane at a much
slower rate than we had anticipated, prevented us from saturating
this lower bound.

It is still interesting to extrapolate our results to obtain
information about the possible behavior of the boundary
in the \hbox{$\epsilon$-$\nu$} plane of topology changing
solutions. For this purpose we binned all our data into
subintervals of width $\Delta \epsilon = 0.005$. Within
every bin we selected the point with lowest $\nu$. We
then fitted these points to the hyperbola
\begin{equation}
 (\nu-\alpha \epsilon-c_1)(\nu-\nu_{\infty})=c_2 \ ,
\label{hypefit}
\end{equation}
where $\alpha$ and $\nu_{\infty}$ are the free parameters
of the fit. The quantities $c_1$ and $c_2$ (which are constants
with respect to $\nu$ and $\epsilon$ but depend on
$\alpha$ and $\nu_{\infty}$) are given by $c_1=2 \nu_{\rm sph}-
\alpha \epsilon_{\rm sph} - \nu_{\infty}$ and $c_2=
-(\nu_{\rm sph}-\nu_{\infty})^2$, where $\epsilon_{\rm sph}=
2.5447$ and $\nu_{\rm sph} = 1.7478$ are the energy and particle
number in the limiting case in which the configuration approaches
the sphaleron itself (in practice the $\epsilon$ and $\nu$ of
the configuration from which we started the Monte Carlo search).

This fit is motivated by simple physical considerations. We
would expect the lower boundary of the region of topology
changing transitions to saturate either at $\nu=0$ or at some
finite value of $\nu$. The boundary of the domain must go
through the sphaleron and should have an infinite slope there.
Indeed, since the topology changing classical solutions become
complex when $\epsilon$ decreases below $\epsilon_{\rm sph}$,
one would expect the boundary curve $\nu=\nu(\epsilon)$ to
have a square root singularity at $\epsilon=\epsilon_{\rm sph}$.
Finally, although the upper boundary of the region is of little
interest to us, it is not unreasonable to parameterize it in terms
of a straight line of constant slope. This is the line one would
find if the upper bound were obtained by putting all the energy
in a single mode of frequency $\omega$ (in which case the slope
$\alpha= 1 / \omega$), or since this is unrealistic, if the the mode
distribution could be well approximated in terms of some effective
frequency $\omega_{\rm eff} = 1 / \alpha$. The hyperbola of
(\ref{hypefit}) is the simplest curve with all these properties.

The results of our fits are shown in Figs.~13~and~14. In Fig.~13
all the data points have been used, and the solid line represents
the unconstrained fit while the dashed line is obtained by
requiring $\nu_\infty=0$. Since one can argue that what ought to
be fit is the lower boundary of the region, and that insofar as
our points display a slight discontinuity and cannot all belong
to this boundary, we have repeated the fit removing all the
points which lie above the unconstrained fit in Fig.~13. The
results of this second fit are reproduced in Fig.~14. For
Fig.~13, the unconstrained fit has  parameters $\alpha=0.257$
and $\nu_{\infty}=-0.294$ while the constrained fit has
$\alpha=0.319$. Figure~14 has the parameters $\alpha=0.238,
\nu_{\infty}=-0.530$ and $\alpha=0.341$ respectively for the
unconstrained and constrained fits.

\vbox{
\begin{figure}
\centerline{
\epsfxsize=100mm
\epsfbox{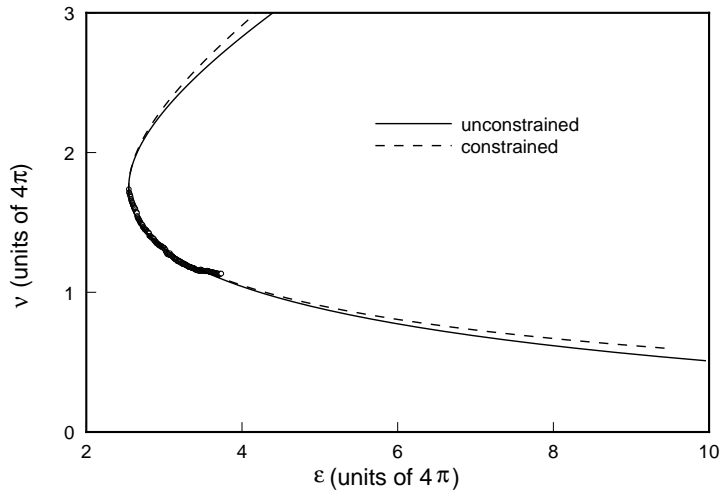}
}
\caption{\tenrm
Hyperbolic fits to full data set. The asymptotic particle
number is constrained to vanish for the dashed line, while
it remains unconstrained for the solid line.
}
\end{figure}
}
\vbox{
\begin{figure}
\centerline{
\epsfxsize=100mm
\epsfbox{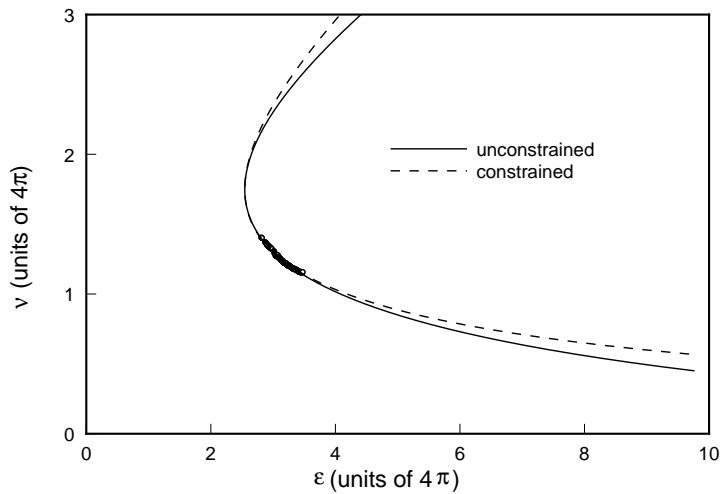}
}
\caption{\tenrm
Same as Fig.~13, except the data set was reduced by
those point lying above the previous unconstrained
fit.
}
\end{figure}
}

\noindent
It is interesting to observe that that the unconstrained fits lead to
an asymptotic value for $\nu$ smaller than zero, which shows that
one cannot read any indication of a lower bound on the particle
number in our present data. Our results cover a range of energies
which is too small to derive any reliable conclusion about whether
and when the particle number could reach the value two. One can
nevertheless insert physical units in the results of our fits and
see at what energy values the incident particle number would
become equal to two. This simple exercise gives energies of
$110.37 {~\rm TeV}$ and $447.20 {~\rm TeV}$ respectively for the
unconstrained and constrained fits of Fig.~13, and energies of
$75.06 {~\rm TeV}$ and $418.61 {~\rm TeV}$ for the corresponding
fits of Fig.~14.

We conclude this section with a few technical remarks.  Since
our entire procedure is based on the calculation of the particle
number after the system has reached the linear regime, we
should make sure this quantity is evaluated in a reliable
manner. Now, it is clear from the graphs of Figs.~7~and~11
that,  while the particle number becomes reasonably constant
towards the end of the evolution, it still exhibits oscillations
possibly as large as 10\%.  This might cast doubts on the
validity of our stochastic sampling technique, where the
steps in initial parameter space induce variations of the
particle number  as small as $10^{-4}$.  The solution
we have adopted is to define a ``computational particle number''
$\nu_c$ (which is the quantity represented in Fig.~9).
With a lattice of infinite spatial extent, even in
presence of an ultraviolet cutoff arising from finite lattice
spacing, and barring the existence of conservation laws
giving rise to particle phenomena, the system will eventually
linearize fully and the true particle number $\nu$ will be well
defined and constant to any degree of precision.  Since we
begin with an initial state localized around the origin, we may
conceptually think of this as being defined over an infinite lattice,
although in practice we use a lattice of finite extent.  Thus,
every initial configuration $c_n$ conceptually determines a
unique  particle number $\nu$.  This may not be accessible to
us, but it exists.  We define a quantity $\nu_c$ which we can
measure as follows:  we evolve the system for a definite amount
of time $T_0$ and then for an additional time $\Delta T$ (in our
calculation  $T_0= 60$ and $\Delta T =8$). Over the interval
$T_0, \, T_0+\Delta T$ we measure the particle number at times
$T_i = T_0 \dots T_m$ chosen at random (in our calculation we
take $m=10$ and choose $T_1$ through $T_{10}$ to be 61.55, 62.51,
63.27, 63.70, 64.77, 65.25, 65.33, 65.71, 66.59, 68.00 respectively), 
but fixed for the entire calculation.  The computational
particle number $\nu_c$ is defined as the average of the particle
numbers measured at $T_i$.  Again, $\nu_c$ is a well defined
function of the parameters $c_n$, and uniquely determined by
this initial configuration.  The crucial point is that $\nu_c$ tracks
$\nu$.  The quantities $\nu$ and $\nu_c$ may differ by as much
as 10\%; however, if we reduce $\nu_c$ by a certain factor, we
can be confident that the true particle number $\nu$ has also
been reduced by the same factor, up to a relative error given
by the approximation by which $\nu_c$ tracks $\nu$. Finally,
we should make sure that $\nu_c$ is, computationally, a well
behaved function  of the parameters $c_n$, i.e.~that the
functional relation between the chosen $c_n$ and the measured
value  of $\nu_c$ is not spoiled by numerical errors.  This we have
verified explicitly.  On a sample configuration we have stepped
every individual parameter $c_n$ by values an order of magnitude
smaller than the typical steps in our stochastic sampling procedure
and have verified that the corresponding changes in $\nu_c$
are regular and well accounted for by the first few terms of a
Taylor series expansion in $\Delta c_n$.

\section{Conclusions}
\hspace{1cm}
We have developed a computational procedure that allows
us to explore the space of classically allowed topology
changing transitions leading to baryon number violation.
With our method we have been able to trace the lower
boundary of the region spanned by topology changing
evolution in the energy versus incoming particle number
plane, up to energies approximately one and a half
times the sphaleron energy and with a reduction of the
incoming particle number by approximately 40\%. The
corresponding solutions display dramatically different
features in their incoming state from the solution used to
seed the Monte Carlo search (in which there was just
barely enough energy to cross the sphaleron barrier),
one of the most notable differences being a marked
shift of the in-state spectral mode distribution towards higher
frequencies and shorter wavelengths. Within the domain
we have explored there is no indication of an emergent lower
limit on the particle number of the incoming state. Indeed,
a hyperbola fit to our data, motivated by the expected physical
properties of the boundary of the domain of topology changing
evolution, is quite compatible with a zero lower bound on the
incoming particle number.

Our results are unfortunately rather limited in the extent
of energy and particle number which we have been able
to explore. However impressive may be the change in the
properties of the solutions spanned by our search, the fact
remains that the lowest particle number we have been able
to reach is, in physical units, approximately 30. An even
more serious shortcoming of our results is that our method
can only establish an upper bound on the minimum particle
number at any given energy: when our search produces a
topology changing solution of given $\epsilon$ and $\nu$,
it establishes by construction that the lower boundary of the
classically allowed transitions cannot lie above that point,
but we cannot rule out that it might lie substantially below
and that the stochastic search simply failed to come close
to it.

However, the mere fact that the analytically intractable
non-linear equations of motion are amenable to a reliable
computational solution is, we believe, a very important result,
perhaps the most important fact emerging from our analysis.
By solution, we mean much more than just the
implementation of a numerical integration algorithm of the
evolution equations. Our study makes it clear that a whole
range of detailed questions about the entire space of solutions
can be tackled and solved by computational means.

The results we have established thus far naturally lead to
further investigation. By investing more computational
resources it will be straightforward to extend the
exploration to substantially larger energies. However, one
can do more than that. The detailed information obtained
about the spectral composition of the incoming states with
low particle number suggests that that one may explore
the properties of such states directly. For instance,
one could try to shift the mode distribution further towards
shorter range, while verifying that the ensuing evolution
still changes topology. This runs somehow against our
original notion that it would be very difficult to start from the
selection of the incoming state and still obtain a topology
changing solution, but now we are no longer dealing
with a blind sampling of incident states. From this point
of view we find very inspiring some recent results obtained
by Farhi, Goldstone, Lue and Rajagopal who, in a study
of collision induced soliton decay, were able to produce
the ``unwinding'' of the soliton and its subsequent decay
by directing against it waves which carry a short range twist
of the phase of the complex field (we refer to the original work
of Ref.~\cite{fglr} for an elucidation of this possibly cryptic
sentence). It is interesting that in computer animation
which we generated to clarify the properties of the evolution,
we have seen analogous twists in the phase of the
\hbox{2-dimensional} field $\chi$ in the asymptotic
states. Of course one must be careful in defining effects
which pertain to gauge variant quantities, but a careful
study of the properties of the asymptotic states may provide
important clues for understanding the mechanisms leading
to classically allowed transitions with low incoming particle
number.

Finally, a complementary approach to the study of classically
allowed transitions consists in studying the classically forbidden
processes. As we have already mentioned in the Introduction,
a very powerful formalism for the study of such processes has
been established in Ref.~\cite{rst} and applied recently in
Ref.~\cite{KuzTin} to the study of collision induced decay of
the false vacuum. The method of Ref.~\cite{rst} requires that
one solves analytically continued equations of motion along
a suitable contour in the complex time plane and that one
imposes boundary conditions based on the normal mode
expansion of the fields in the linearized domain. Thus a
large part of the formalism we have developed in this
paper will carry over to the study of classically forbidden
processes, and we plan to make this the subject of a future
investigation.

{\bf Acknowledgments}

This research was supported
in part under DOE grant DE-FG02-91ER40676 and NSF grant
ASC-940031. We wish to thank V.~Rubakov for very interesting
conversations which stimulated the investigation described here,
A.~Cohen, K.~Rajagopal and P.~Tinyakov for valuable discussions,
and T.~Vaughan for participating in an early stage of this work.

\end{document}